%% file: nsdi24.tex
\documentclass[letterpaper,twocolumn,10pt]{article}
\usepackage{usenix20}
\input{packages}

\newcommand{\DC}{\texttt{Acme}\xspace}
\newcommand{\Scluster}{\texttt{Seren}\xspace}
\newcommand{\Kcluster}{\texttt{Kalos}\xspace}

\begin{document}

\date{}

\newcommand{\papername}{Characterization of Large Language Model Development in the Datacenter}

\title{\Large \bf \papername}

\author{
Qinghao Hu\textsuperscript{\footnotemark[1] \ \ding{73}1}, Zhisheng Ye\textsuperscript{\footnotemark[1] \ \ding{73}3}, Zerui Wang\textsuperscript{\thanks{Equal Contribution.} \ \ding{73}4}, Guoteng Wang\textsuperscript{\ding{73}}, Meng Zhang\textsuperscript{\ding{73}1}, Qiaoling Chen\textsuperscript{\ding{73}1}
\\
Peng Sun\textsuperscript{\ding{73}5}, Dahua Lin\textsuperscript{\ding{73}6}, Xiaolin Wang\textsuperscript{3}, Yingwei Luo\textsuperscript{3}, Yonggang Wen\textsuperscript{2}, Tianwei Zhang\textsuperscript{2}
\vspace{5pt}
\\
{\it \textsuperscript{\ding{73}}Shanghai AI Laboratory}
\qquad
{\it \textsuperscript{1}S-Lab, Nanyang Technological University}
\qquad
{\it \textsuperscript{2}NTU}
\\
{\it \textsuperscript{3}Peking University}
\qquad
{\it \textsuperscript{4}Shanghai Jiao Tong University}
\qquad
{\it \textsuperscript{5}SenseTime Research}
\qquad
{\it \textsuperscript{6}CUHK} 
}

\maketitle

\input{0_Abstract}
\input{1_Introduction}
\input{2_Background}
\input{3_Datacenter}

\input{4_Profiling}

\input{5_Failure}
\input{6_System}
\input{7_Related}

\input{8_Discussion}
\input{9_Conclusion}

\section*{Acknowledgments}
We sincerely thank our shepherd, Guyue Liu, and anonymous NSDI reviewers for their valuable comments on this paper.
We thank Penglong Jiao, Wenwei Zhang, Xingpu Li, and the broader InternLM team for their support throughout this project.
The research is supported under the National Key R\&D Program of China (2022ZD0160201) and the RIE2020 Industry Alignment Fund - Industry Collaboration Projects (IAF-ICP) Funding Initiative, as well as cash and in-kind contributions from the industry partner(s).

\bibliographystyle{plain}
\bibliography{nsdi24}

\input{10_Appendix}

\end{document}

%% file: packages.tex
\usepackage{xcolor}
\usepackage{xspace}
\usepackage{mathtools} 
\usepackage{amssymb}
\usepackage{bm}
\usepackage{enumitem}
\usepackage{textcomp}

\usepackage{graphicx}
\usepackage{subcaption}
\usepackage{tikz}
\usepackage[labelfont=bf]{caption}

\usepackage{tabularx}
\usepackage{multirow}
\usepackage{booktabs}
\usepackage{makecell}
\usepackage[figuresleft]{rotating}
\usepackage[flushleft]{threeparttable}
\usepackage{colortbl}

\usepackage{algorithmicx}
\usepackage{algorithm}
\usepackage[noend]{algpseudocode}

\algnewcommand\algorithmicinput{\textbf{Input:}}
\algnewcommand\Input{\item[\algorithmicinput]}
\algnewcommand\algorithmicoutput{\textbf{Output:}}
\algnewcommand\Output{\item[\algorithmicoutput]}

\usepackage[utf8]{inputenc} 
\usepackage{inconsolata} 
\usepackage[most]{tcolorbox} 
\usepackage{soul} 
\usepackage{siunitx} 
\usepackage{pifont} 
\usepackage{xurl} 
\usepackage[textsize=scriptsize, linecolor=magenta, bordercolor=magenta,
  backgroundcolor=white, textwidth=45pt]{todonotes}
\usepackage{marginnote}
\usepackage[noindentafter]{titlesec}
\titlespacing\section{0pt}{3pt}{3pt} 
\titlespacing\subsection{0pt}{3pt}{3pt}
\titlespacing\subsubsection{0pt}{5pt}{5pt}

\usepackage[skip=3pt]{caption}


%% file: 0_Abstract.tex
\begin{abstract}
  Large Language Models (LLMs) have presented impressive performance across several transformative tasks. However, it is non-trivial to efficiently utilize large-scale cluster resources to develop LLMs, often riddled with numerous challenges such as frequent hardware failures, intricate parallelization strategies, and imbalanced resource utilization. In this paper, we present an in-depth characterization study of a six-month LLM development workload trace collected from our GPU datacenter \DC. Specifically, we investigate discrepancies between LLMs and prior task-specific Deep Learning (DL) workloads, explore resource utilization patterns, and identify the impact of various job failures.
  Our analysis summarizes hurdles we encountered and uncovers potential opportunities to optimize systems tailored for LLMs. Furthermore, we introduce our system efforts: (1) \emph{fault-tolerant pretraining}, which enhances fault tolerance through LLM-involved failure diagnosis and automatic recovery. (2) \emph{decoupled scheduling for evaluation}, which achieves timely performance feedback via trial decomposition and scheduling optimization.


\end{abstract}

%% file: 1_Introduction.tex
\section{Introduction}
\label{sec_intro}
Over the years, advances in LLMs have attracted significant attention from both academia and industry owing to their impressive performance and capabilities, such as ChatGPT \cite{ChatGPT} and GitHub Copilot \cite{GitHubCopilot}. However, due to their immense model sizes and extensive data demands, training such models necessitates a substantial computational infrastructure with thousands of accelerators \cite{MegatronLM, PaLM}. Hence, it is a common practice for tech companies and cloud providers to build large-scale GPU clusters to facilitate LLM development, especially after the popularity of ChatGPT.
Nevertheless, it is non-trivial to perform efficient LLM development on such high-cost infrastructure. Developers often confront numerous issues and challenges, including frequent hardware failures \cite{GEMINI, CheckFreq}, intricate parallelization strategies \cite{Alpa, MegatronLM}, unstable training progress \cite{OPT, BLOOM}, long queuing delay \cite{Orca}, etc.

Developing LLMs is closely intertwined with the support of GPU clusters in various aspects. A thorough analysis of cluster workloads is essential for comprehending challenges and uncovering opportunities in designing systems tailored for LLMs. However, many conclusions and implications from existing DL workloads analysis works \cite{Philly,Helios,MLaaS}, conducted before the rise of LLMs, are not applicable to LLM development. This is primarily due to the divergent characteristics and requirements of LLMs:

\noindent\textbf{(1) \emph{Paradigm Transition}}.
DL workloads generally follow a \emph{task-specific} paradigm that trains the model on domain-specific data to tackle a particular task (e.g., translation \cite{LSTM}). In contrast, LLMs follow an emerging paradigm that performs self-supervised training on broad data to generate a \emph{foundation model} \cite{FoundationModel} and further adapts to a wide range of downstream tasks. This shift signifies a substantial divergence in the model development pipeline (e.g., pretraining \cite{Megatron-LMV1}, alignment \cite{LoRA}) and workload characteristics from prior DL workloads (\S \ref{sub_sec_llm_pipline}).

\noindent\textbf{(2) \emph{Tailored Software Stack}}.
To accommodate the enormous model size of LLMs, a series of systems implement advanced techniques to optimize the execution of LLMs.
For instance, Deepspeed \cite{ZeRO}, Megatron \cite{MegatronLM} and Alpa \cite{Alpa} accelerate the training via hybrid parallelism or state-sharding optimizer. As for model serving, Orca \cite{Orca} and vLLM \cite{vLLM} improve throughput via iteration scheduling or memory management.

\noindent\textbf{(3) \emph{Unified Architecture}}.
Prior DL workloads usually employ various model architectures (e.g., CNN \cite{CNN}, RNN \cite{LSTM}) to address diverse tasks. In contrast, LLMs commonly embrace the Transformer \cite{Transformer} architecture, like BERT \cite{BERT}, GPT-3 \cite{GPT-3}, LLaMA \cite{LLaMA} and PaLM \cite{PaLM}.
The architectural homogeneity suggests a high level of uniformity in the LLM development pipeline and similarity across different datacenters.

To bridge this gap, we present an in-depth study of our operational experiences in the datacenter \DC of Shanghai AI Laboratory. It houses two distinct clusters, \Scluster and \Kcluster, dedicated to LLM development and equipped with 4,704 A100 GPUs in total. Our analysis draws upon traces collected over a six-month period from March to August 2023, encompassing scheduler logs, infrastructure monitoring data, failure logs, and fine-grained profiling data.
Our key findings and identified challenges can be summarized as follows:

\begin{itemize}[leftmargin=*,topsep=0pt, itemsep=-3pt, itemindent=8pt]
      \item \textrm{\bfseries Shorter Job Duration and Unfair Queuing Delay}.
            In contrast to the common stereotype that LLM workloads are usually long-term, the workloads in our datacenter exhibit 2.7$\sim$12.8$\times$ shorter average job duration compared to the DL workloads in previous traces \cite{Philly, Helios, MLaaS}.
            This can be attributed to the presence of numerous short-term tasks such as evaluation. 
            In terms of job queuing delay, our findings also diverge from previous DL traces that larger-scale jobs experience longer wait times. We observe that evaluation jobs, despite being short-term and small-scale, have the longest queuing delay. This discrepancy stems from reserving the majority of resources for pretraining jobs to minimize their queuing delays. Evaluation jobs are scheduled with a lower priority, utilizing the limited spare resources.

      \item \textrm{\bfseries Imbalanced Resource Usage}.
            The imbalance is manifested in two aspects. Firstly, in terms of workload distribution, pretraining jobs only account for 3.2\% of the total job count but consume 94.0\% of the whole compute resource (i.e., GPU time) in \Kcluster. Conversely, evaluation jobs, despite constituting 92.9\% of all jobs, only utilize a meager 0.8\% of resources.
            Secondly, when looking at infrastructure utilization, we find that associated resources including CPU, host memory, and network, are frequently underutilized. In contrast, the GPU, as the primary resource, shows high utilization rates. Both GPU memory and GPU utilization exhibit substantially higher median values at 75\% (60GB) and 99\% respectively in \Kcluster, as opposed to the 10\% and 4\% observed in PAI \cite{MLaaS}.
            These observations corroborate that LLMs are computationally and memory intensive. It also implies that GPU-sharing-based techniques \cite{Gandiva, AntMan, Lucid, Salus} may not be suitable for LLMs.

      \item \textrm{\bfseries Long GPU Idle Time in Evaluation Workload}.
            Our profiling of evaluation workloads reveals substantial under-utilization of GPU resources at various stages. For example, the evaluation job on HumanEval consumes 29.5\% of its time for model loading and data preprocessing, and an additional 19.0\% is spent conducting synthesized program correctness tests. As a result, only half of the time is dedicated to GPU inference, leading to long queuing delays in evaluation trials.

      \item \textrm{\bfseries Frequent Job Failures}.
            We find various errors primarily occur at the beginning of LLM workloads, leading to fast job termination. However, infrastructure failures, which are common in long-term pretraining jobs, significantly impede training efficiency. Therefore, prompt diagnosis and recovery from these failures are crucial to enhance training efficiency.
            
\end{itemize}

Based on our characterization study, we identify several challenges encountered during the LLM development, such as unstable training progress, remote storage bottleneck and delayed feedback on model performance. To tackle these issues, we consolidate the insights gained from our operational experience and build two systems that are integrated into our LLM framework to improve development robustness and efficiency. \emph{Firstly}, to mitigate the frequent failure problem, we establish a system to achieve \textbf{\emph{fault-tolerant pretraining}}. It incorporates three key designs: (1) achieving frequent model saving through asynchronous checkpointing, (2) identifying the root causes of various failures through a combination of heuristic rules and LLM, (3) employing a holistic detection toolkit to pinpoint fault nodes and automatically restart training from properly saved checkpoint. It accelerates checkpointing by 3.6$\sim$58.7$\times$ and significantly reduces manual intervention.
\emph{Secondly}, we develop a system that performs \textbf{\emph{decoupled scheduling for evaluation}} to provide developers with timely feedback on model quality. It not only resolves the remote model loading contention issue via decoupled model retrieval but also minimizes GPU idle time via decoupling the metric computation process. It further leverages the prior knowledge and flexibility of datasets to balance workload across all GPUs. Our experiment shows that it can reduce the evaluation makespan by up to 1.8$\times$.

We believe the observations and insights derived from our datacenter do not stand in isolation. Our traces are publicly available at \url{https://github.com/InternLM/AcmeTrace}. We also release our system code and models (Appendix \ref{appendix_artifact}). We hope these resources and lessons can benefit researchers in optimizing LLM systems as well as GPU cluster management.


%% file: 2_Background.tex
\section{Background}
\label{sec_background}

\begin{figure}[t]
    \centering
    \includegraphics[width=\linewidth]{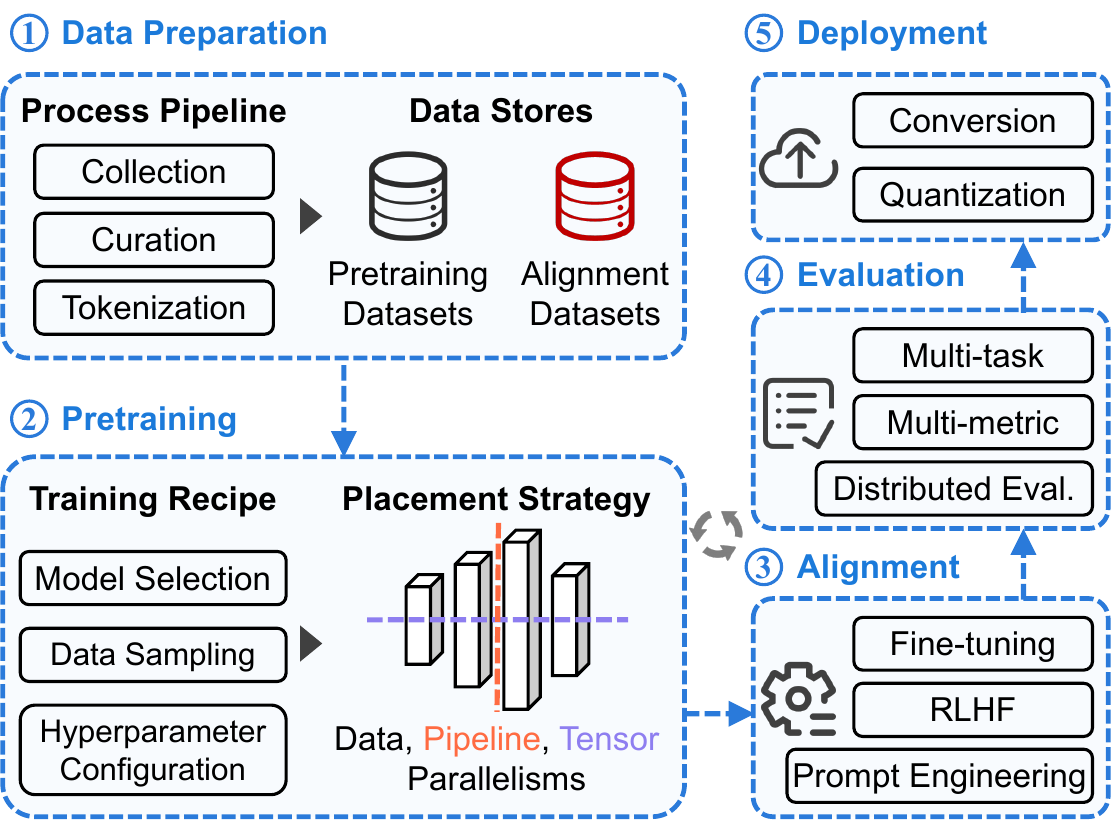}
    \caption{Overview of the LLM development pipeline.}
    \label{figure_llm_pipeline}
\end{figure}

\subsection{LLM Development Pipeline}
\label{sub_sec_llm_pipline}
Distinguished from task-specific DL models, LLMs follow an emerging paradigm that performs self-supervised training on broad data and further adapts to a wide range of downstream tasks \cite{FoundationModel}. The development of LLMs necessitates the use of extensive computational infrastructure due to their substantial model size (comprising billions of parameters) and the vast amount of training data involved.
Figure \ref{figure_llm_pipeline} depicts the comprehensive LLM development pipeline, encompassing five distinct stages (blue blocks) that span from scratch to service (follow blue arrows). The grey circular arrow indicates that the pretraining stage enables periodical alignment and evaluation to assess intermediate models and adjust configuration on the fly. We explain each stage as follows:

\noindent\textbf{Data Preparation}.
The initial stage involves gathering and preprocessing the training data, which can be categorized into two parts: (1) \emph{pretraining data}, consisting of extensive unlabeled corpora obtained from public or private sources and curated through processes like detoxification and deduplication; (2) \emph{alignment data}, comprising a smaller set of high-quality labeled corpora used to align the model with specific tasks. This data is typically acquired through expensive human annotation or labeling \cite{InstructGPT}. Besides, all the data must be tokenized to ensure compatibility with the model's input.

\noindent\textbf{Pretraining}.
It involves self-supervised training on large-scale curated data, demanding a majority of resources within the overall development workflow. Training LLMs efficiently at scale necessitates various system innovations, such as state-sharding optimizers \cite{ZeRO}, meticulous model placement using data, pipeline, and tensor parallelisms \cite{PipeDream, MegatronLM, Alpa}.


\noindent\textbf{Alignment}.
This stage aims to adapt LLMs with user intent on a wide range of downstream tasks. Two primary aligning paradigms are commonly used:
(1) \emph{prompt engineering}, specifying prompts (i.e., inputs) without modifying model parameters. For example, in text summarization, appending a prompt \textsl{TL; DR} to the input article can improve model performance \cite{GPT-2}; (2) \emph{fine-tuning}, updating model parameters on a task-specific dataset to improve performance in a particular domain.  Additionally, reinforcement learning from human feedback (RLHF) \cite{InstructGPT} further enhances the alignment effect, and parameter-efficient techniques like LoRA \cite{LoRA} have been proposed to reduce the cost of fine-tuning.

\noindent\textbf{Evaluation}.
Given the vast application scenarios of LLM, it may be inaccurate to assess model quality solely based on a single metric like training loss. There are numerous factors to consider, such as accuracy, fairness, and toxicity \cite{HELM}. Consequently, it is crucial to account for a diverse set of criteria and measure performance across multiple tasks \cite{LLMEvalSurvey}. Furthermore, regular evaluation is essential during the pretraining stage to provide timely feedback on model quality.

\noindent\textbf{Deployment}.
To meet the strict cost and latency constraints of LLM applications, several advanced techniques have been developed to achieve efficient model serving, including quantization \cite{LLMint8}, distillation \cite{DistilBERT}, CUDA kernel optimization \cite{FlashAttention, DataMovement}, model parallelism \cite{AlpaServe, Orca} and memory management \cite{vLLM}.

\begin{table}[t]
    \centering
    \renewcommand{\arraystretch}{1.25}
    \resizebox{\linewidth}{!}{
        \begin{tabular}{@{}lcccccc@{}}
            \toprule
            Cluster        & \#CPUs               & \#GPUs             & Mem(GB) & Network          & \#Nodes \\ \midrule
            \textbf{Seren} & \multirow{2}{*}{128} & \multirow{2}{*}{8} & 1,024   & 1$\times$200Gb/s & 286     \\
            \textbf{Kalos} &                      &                    & 2,048   & 5$\times$200Gb/s & 302     \\ \bottomrule
        \end{tabular}
    }
    \caption{Summary of per-node specification and cluster scale for two  independent LLM clusters in \DC.}
    \label{table_cluster_summary}
\end{table}

\subsection{\DC Overview}
\DC is our private GPU datacenter that empowers researchers and engineers to develop DL models across diverse domains. In this work, we focus on analyzing workloads within two clusters dedicated to developing LLMs: \Scluster and \Kcluster. We collect and analyze all jobs in these two clusters. Note that there are additional clusters within \DC that are designated for different fields, such as autonomous driving, and AI for scientific research. However, these clusters are excluded in this work as they are unrelated.

\noindent \textbf{Cluster Architecture}.
Table \ref{table_cluster_summary} summarizes configurations of these two homogeneous LLM clusters. \Scluster and \Kcluster have 2,288 and 2,416 GPUs respectively. Each node is equipped with 8$\times$ NVIDIA A100-SXM 80GB GPUs \cite{A100} and 2$\times$ Intel Xeon Platinum 8358P CPUs (128 threads in total). GPUs are interconnected to each other by NVLink and NVSwitch \cite{nvlink}, and inter-node communication is achieved via NVIDIA Mellanox 200Gbps HDR InfiniBand \cite{InfiniBand}. Compared to \Scluster, \Kcluster is a relatively newer cluster with an improved network configuration. Each node in the \Kcluster has a larger host memory (2TB) and is equipped with four InfiniBand HCAs specifically for application communication, along with an extra HCA dedicated to storage.

Besides, the distributed storage system is also critical for workload performance. \DC adopts an all-NVMe shared parallel file system for fast data access and storage. Moreover, as time has advanced, our resource scheduling system has evolved to support diverse cluster environments. Specifically, our scheduler on \Scluster and \Kcluster is built atop Slurm \cite{SLURM} and Kubernetes \cite{BorgK8S} respectively. In order to provide resource guarantees for large-scale pretraining jobs, our scheduler enables resource isolation and quota reservation. It further incorporates a best-effort job mechanism for higher utilization.

\noindent \textbf{LLM Workloads}.
We develop a collection of LLMs\footnote{Model: \url{https://huggingface.co/internlm}} ranging from 7B to over 123B parameters. All of these models follow the transformer-based decoder-only architecture, similar to the GPT \cite{GPT-1, GPT-2, GPT-3} and LLaMA \cite{LLaMA, LLaMA2} series.  \DC encompasses tasks in the aforementioned general LLM development pipeline (\S \ref{sub_sec_llm_pipline}). Note that \DC does not involve any serving jobs, as our LLMs are deployed on a separate cluster specifically for serving purposes.

\noindent \textbf{Software Stack}.
To support the training of billion-scale models across thousands of GPUs, we built a system InternEvo\footnote{System: \url{https://github.com/InternLM/InternEvo}}, which integrates various system optimization techniques, such as FlashAttention \cite{FlashAttention, FlashAttention2}, 3D parallelism \cite{MegatronLM}, zero redundancy optimization \cite{ZeRO}, mixed precision training \cite{AMP}, selective activation recomputation \cite{Megatron-LMV3} and fine-grained communication overlap. Moreover, it accommodates additional tasks such as model fine-tuning and evaluation.

\subsection{Traces from \DC}
The optimization of LLM-tailored systems and datacenter management can significantly boost development efficiency and yield substantial financial benefits. Achieving this goal requires a profound understanding of the intrinsic characteristics of LLM workloads. Many insights in existing DL workloads analysis works \cite{Philly, Helios, MLaaS} are not applicable to LLM workloads due to the unique attributes of LLMs.
To fill this gap, we collected and analyzed workload traces from our datacenter \DC. Table \ref{table_trace_compare} compares the specifications and trace information of \DC with prior trace analysis works conducted by Microsoft, SenseTime, and Alibaba. Unlike \DC, which is solely dedicated to LLM development, these datacenters encompass a mixture of general DL workloads from various domains. For instance, Helios \cite{Helios} consists of 4 clusters dedicated to training models in computer vision and reinforcement learning, while PAI \cite{MLaaS} includes a diverse range of servers for training and serving jobs.


\noindent \textbf{Trace Source}.
Our characterization study is based on traces collected from two LLM clusters in \DC.
The traces span 6 months from March to August 2023.
\Scluster contains 368K CPU jobs and 664K GPU jobs, while \Kcluster job trace consists of 42K CPU jobs and 20K GPU jobs. Additionally,  we provide a summary of the data sources for the traces used in our study: (1) \emph{Job Log}. We collect the job logs from our scheduler database, which contains detailed information for each job. This includes the job's execution time (submission, start, and end), final status (completed, canceled, failed), requested resources (CPU, GPU, memory), work directory, and other relevant data. (2) \emph{Hardware Monitor Data}. This encompasses long-term, multi-dimensional data obtained from various sources. We collect CPU, memory, and network usage data from Prometheus \cite{Prometheus} database, GPU-related metrics from NVIDIA DCGM \cite{nvdcgm}, and power-related data from IPMI \cite{ipmi}. The sampling interval for this data is set at 15 seconds. (3) \emph{Runtime Log}. To conduct a precise job failure analysis, we capture stdout and stderr logs from LLM frameworks during job execution. (4) \emph{Profiling Data}. For a subset of representative jobs, we delve deeper by performing fine-grained profiling using tools like DCGM. The synergy of these trace dimensions allows us to gain a holistic understanding of LLM job characteristics in datacenters.


\begin{table}[t]
    \vspace{-10pt}
    \centering
    \renewcommand{\arraystretch}{1.25}
    \resizebox{\linewidth}{!}{
        \begin{tabular}{lccccc}
            \toprule
                         & \multicolumn{3}{c}{For Task-Specific DL Models} & \multicolumn{2}{c}{For LLMs}                                               \\ \cmidrule(lr){2-4} \cmidrule(lr){5-6}
            Datacenter   & \textbf{Philly} \cite{Philly}                   & \textbf{Helios}  \cite{Helios} & \textbf{PAI} \cite{MLaaS} & \textbf{Acme} \\ \midrule
            Year         & 2017                                            & 2020                           & 2020                      & 2023          \\
            Duration     & 3 months                                        & 6 months                       & 2 months                  & 6 months      \\
            \#Jobs       & 113K                                            & 3.36M                          & 1.26M                     & 1.09M         \\
            Avg. \#GPUs  & 1.9                                             & 3.7                            & 0.7                       & 6.3           \\
            GPU Model    & 12GB/24GB                                       & 1080Ti/V100                    & T4/P100/V100              & A100          \\
            Total \#GPUs & 2,490                                           & 6,416                          & 6,742                     & 4,704         \\

            \bottomrule
        \end{tabular}
    }
    \caption{Comparison between our datacenter \DC and GPU datacenters in prior trace analysis works: Microsoft Philly \cite{Philly}, SenseTime Helios \cite{Helios}, Alibaba PAI \cite{MLaaS}. Philly only provides GPU memory sizes without clarifying GPU models. The average number of requested GPUs in PAI can be less than 1 (0.7), as it supports fractional (<1) GPU requests.}
    \label{table_trace_compare}
\end{table}

%% file: 3_Datacenter.tex
\section{Datacenter Characterization}
\label{sec_characterization}

In this section, we perform a thorough analysis of \DC, including comparing workload distribution between LLMs and previous DL workloads (\S \ref{sub_sec_compare}), investigating different LLM workload types (\S \ref{sub_sec_workload_analysis}), exploring resource utilization patterns (\S \ref{sub_sec_infrastructure_analysis}) and assessing environmental impacts (\S \ref{sub_sec_environmental}).

\begin{figure}[t]
    \vspace{-10pt}
    \centering
    \includegraphics[width=\linewidth]{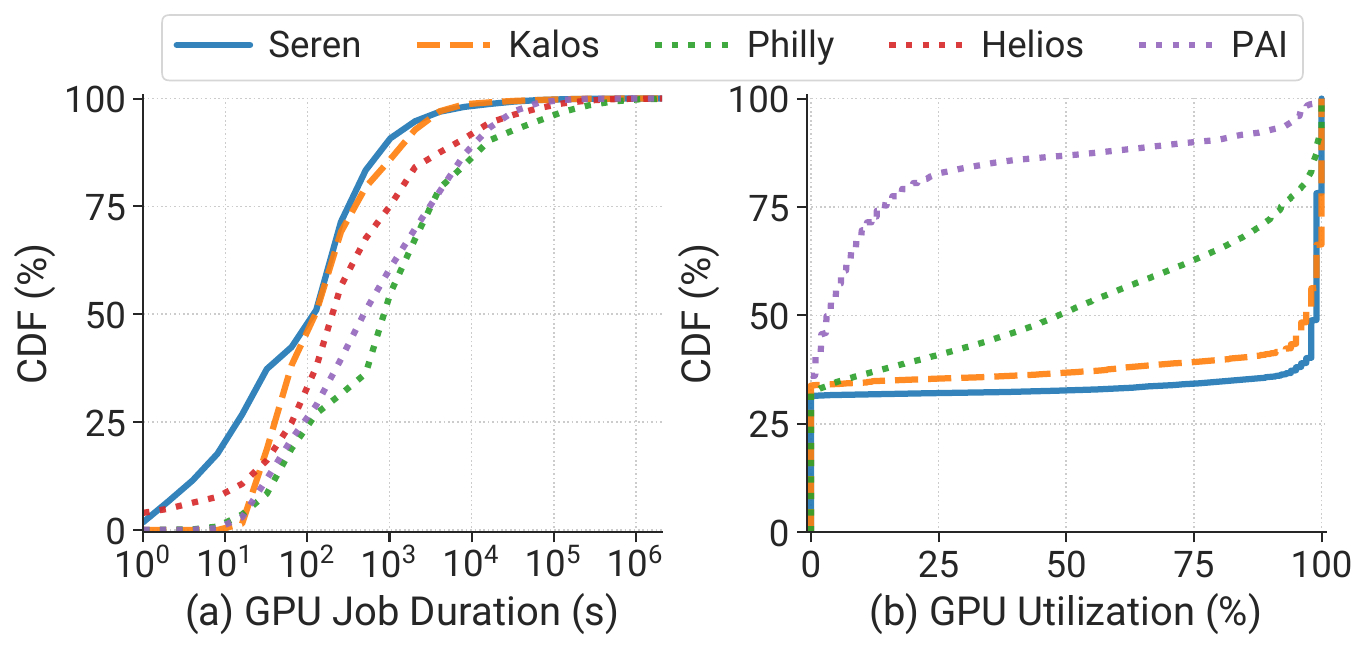}
    \caption{Overview of different datacenter characteristics. (a) \emph{Workload}: CDF of the GPU job duration. (b) \emph{Infrastructure}: CDF of GPU utilization, where Helios' data is not available.}
    \label{figure_compare_overview}
\end{figure}

\begin{figure}[t]
    \vspace{-5pt}
    \centering
    \includegraphics[width=\linewidth]{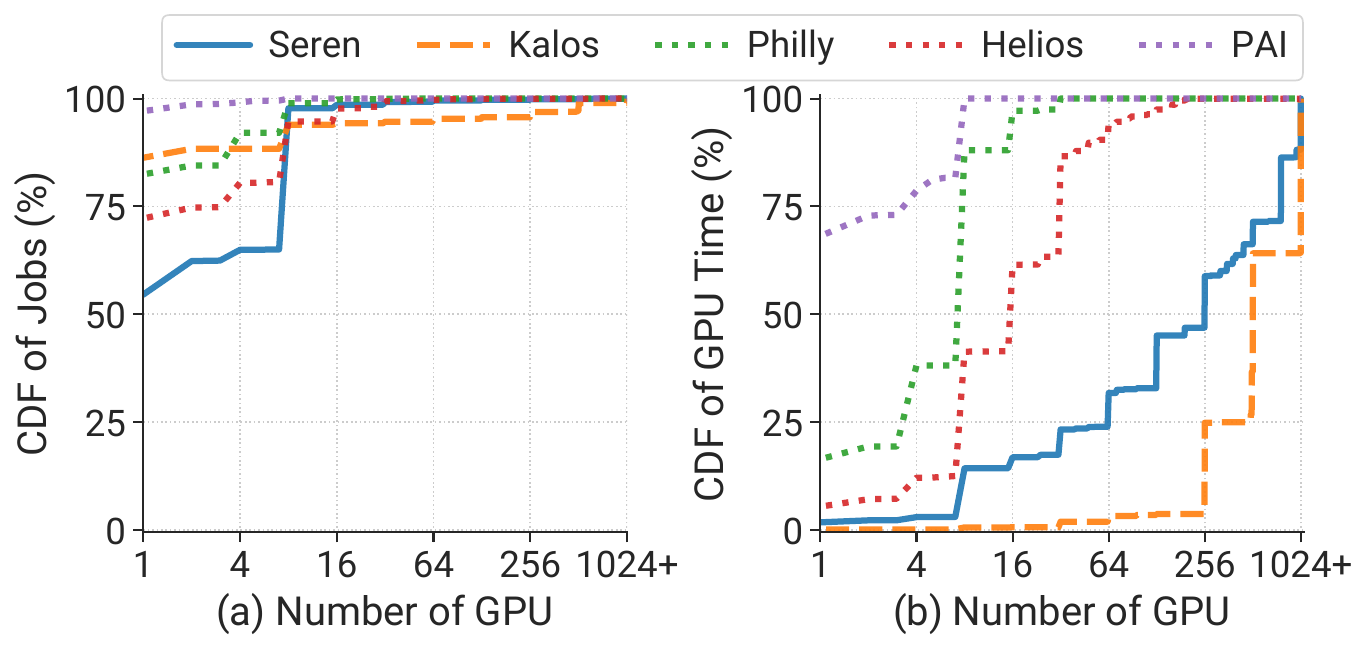}
    \caption{Comparison of workload distribution based on the number of requested GPUs. (a) CDF of job count. (b) CDF of GPU time (i.e., requested GPU number $\times$ duration).}
    \label{figure_compare_gpunum}
\end{figure}

\subsection{LLMs versus Prior DL Workloads}
\label{sub_sec_compare}

\noindent\textbf{Shorter Job Duration}.
As shown in Figure \ref{figure_compare_overview} (a), contrary to the prevailing stereotype that LLM-related jobs are typically long-running, we find the workloads in our clusters (blue and orange lines) exhibit shorter GPU job durations (i.e., job runtime, excluding queuing delay) compared to the DL workloads observed in previous job traces (dotted lines). Specifically, both the \Scluster and \Kcluster have a median job duration of 2 minutes, which is 1.7$\sim$7.2$\times$ shorter than the median job durations of other clusters.
Furthermore, it is evident that the more recent trace demonstrates a shorter job duration distribution. In particular, when considering the average job duration in the Philly cluster (collected in 2017), it is 2.7$\sim$3.8$\times$  longer than Helios (2020) and PAI (2020), and 12.8$\times$ longer than \DC (2023).
To provide an explanation for this observation, we outline four potential factors:
(1) \emph{Hardware upgrade}. The iteration of GPU and networking delivers substantial efficiency improvement.
(2) \emph{Abundant resources}. Users usually request more resources (as shown in Table \ref{table_trace_compare}), averaging 5.7 GPUs in the \Scluster and 26.8 GPUs in the \Kcluster. This can significantly accelerate the training process.
(3) \emph{Extensive associated workloads}: LLM development pipeline involves numerous small-scale associated jobs, such as evaluation. We will delve into this in \S \ref{sub_sec_workload_analysis}.
(4) \emph{High incompletion rate}: Approximately 40\% of jobs fail, with completed jobs consuming only 20$\sim$30\% of GPU resources. This highlights the urgent need for a fault-tolerant system. Further details can be found in Figure \ref{figure_job_state} and Appendix \ref{appendix_subsec_final_statuses}.

\begin{figure}[t]
    \vspace{-10pt}
    \centering
    \includegraphics[width=\linewidth]{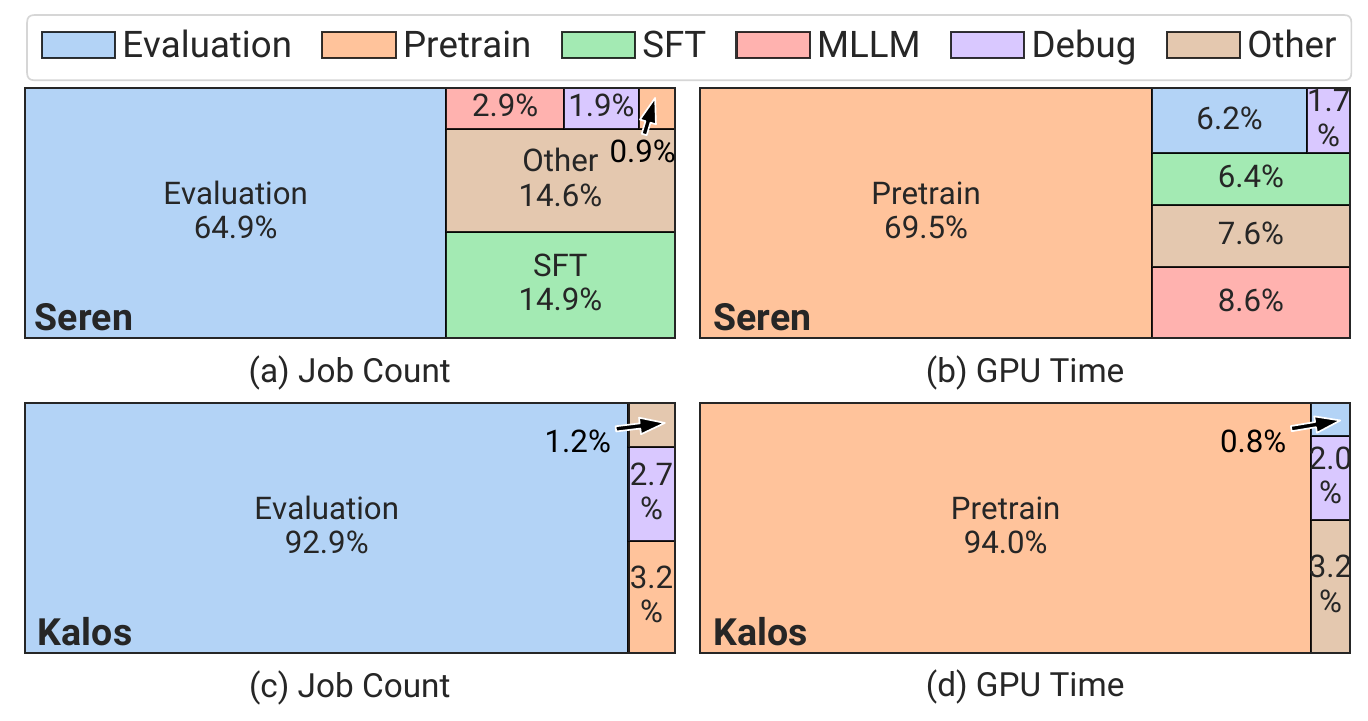}
    \caption{Distribution of different workload types in \Scluster (a, b) and \Kcluster (c, d). Note that CPU jobs are excluded. \emph{SFT}: Supervised Fine-Tuning for model alignment. \emph{MLLM}: Multimodal Large Language Model. \textsl{Other}: Unclassified jobs.}
    \label{figure_treemap_job_types}
\end{figure}

\begin{figure}[t]
    \vspace{-5pt}
    \centering
    \includegraphics[width=\linewidth]{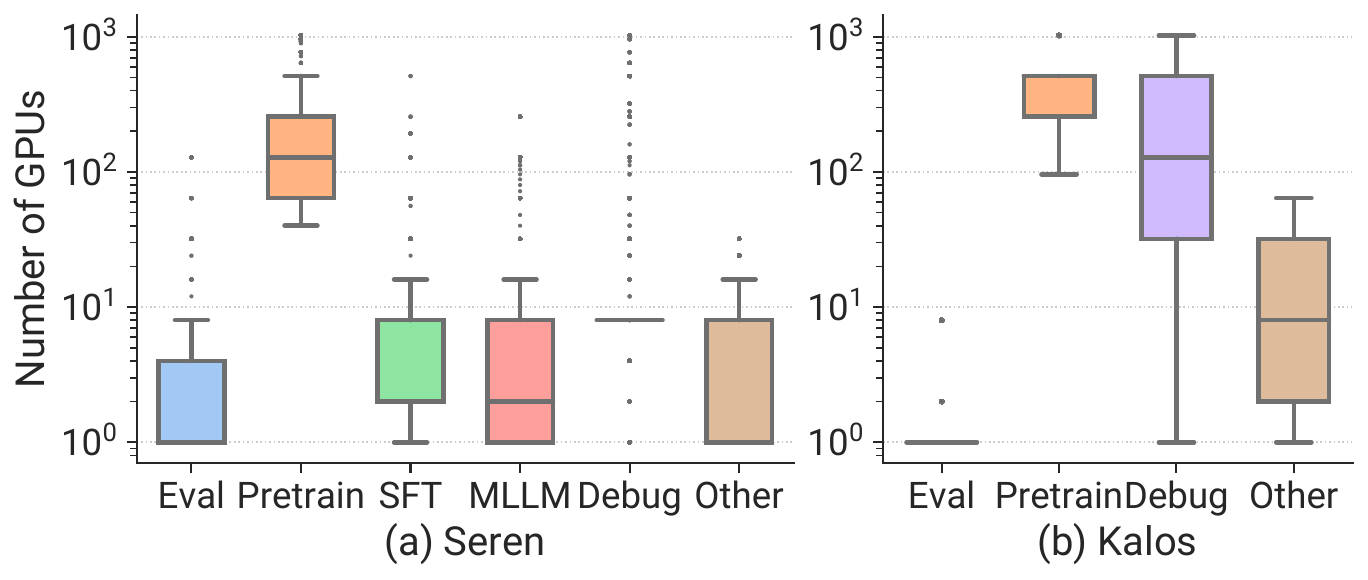}
    \caption{The boxplot of the distribution of GPU demand across different workload types in \Scluster (a)  and \Kcluster (b).}
    \label{figure_boxplot_job_types}
\end{figure}

\noindent\textbf{Polarized GPU Utilization}.
Figure \ref{figure_compare_overview} (b) shows cluster-wide GPU utilization distributions across various clusters. It is evident that the GPU utilization in our two clusters exhibits a polarized pattern, primarily concentrated in two distinct states: 0\% and 100\%.
This polarization mainly stems from the fact that the workloads in our clusters share similar model architectures, i.e., transformer-based LLMs.
In contrast, Philly and PAI encompass a broader range of utilization. Besides, when comparing the median GPU utilization, \Scluster and \Kcluster exhibit significantly higher values at 97\% and 99\%, respectively, in contrast to 48\% and 4\% observed in Philly and PAI.
This observation aligns with the common understanding that LLMs are computationally intensive. It also implies that GPU-sharing-based scheduling techniques \cite{Gandiva, AntMan, Lucid, Salus} may not be suitable for LLM development. Note that `GPU utilization' may sometimes be a weak utilization indicator \cite{nvsmi, HFTA}. We provide a more precise utilization analysis in \S \ref{sub_sec_infrastructure_analysis}.

\begin{figure}[t]
    \vspace{-10pt}
    \centering
    \includegraphics[width=\linewidth]{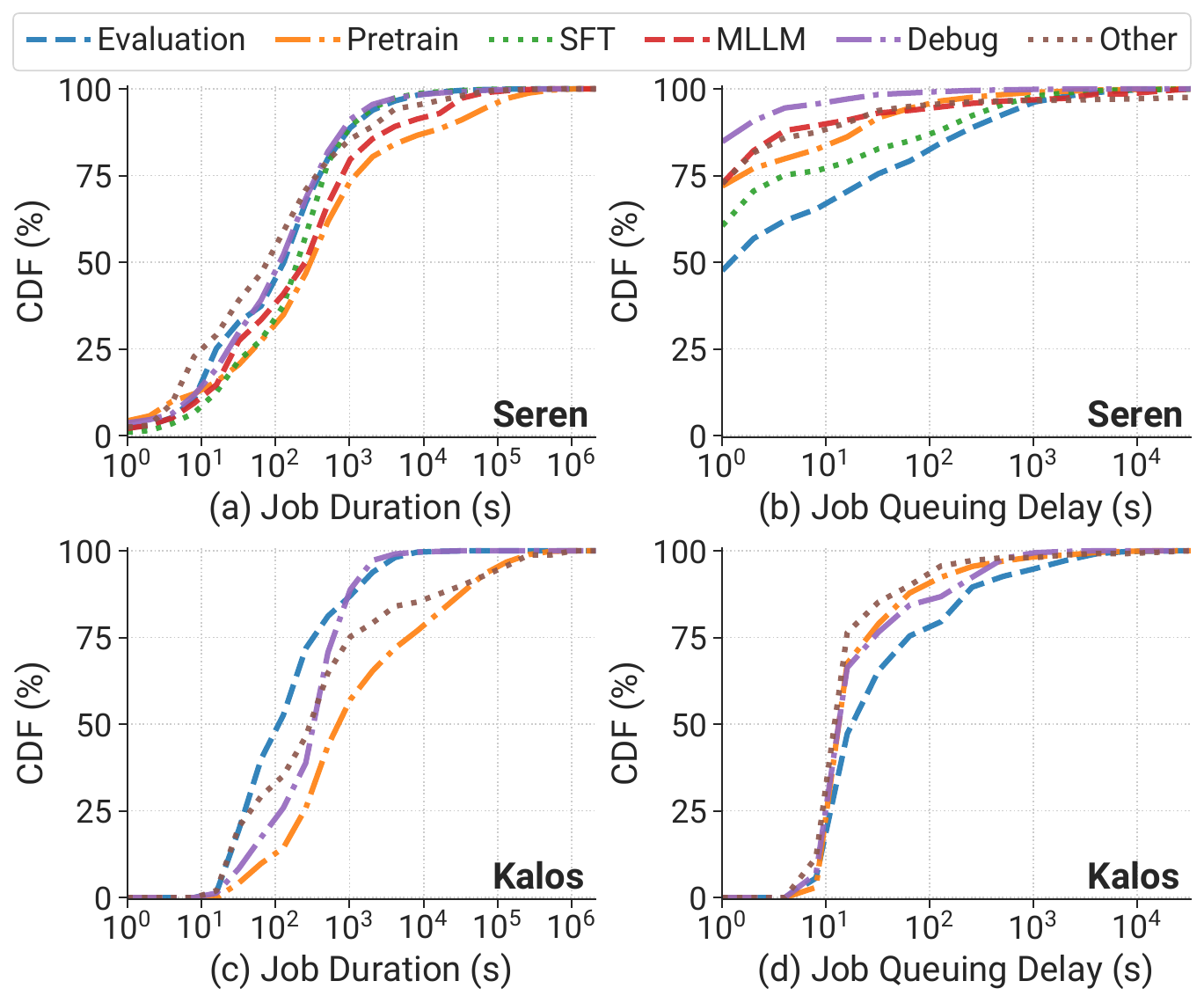}
    \caption{CDF of GPU job duration and queuing delay for different workload types in \Scluster (a, b) and \Kcluster (c, d).}
    \label{figure_cdf_job_types}
\end{figure}

\noindent\textbf{High-skewed Workload Distribution}.
We further investigate the CDF of GPU demands in relation to the number of jobs (Figure \ref{figure_compare_gpunum} (a)) and GPU time (Figure \ref{figure_compare_gpunum} (b)). For the number of jobs, all the clusters share a similar pattern in that the majority of jobs are single-GPU jobs and less than 7\% of jobs request over 8 GPUs. However, when examining GPU time, single-GPU jobs only account for less than 2\% resources in our two clusters, while taking over 68\% GPU time in PAI. In stark contrast, large-scale jobs ($\geq$ 256 GPUs) dominated the GPU time in \Kcluster, occupying more than 96\% of resources.
The much steeper distribution poses substantial challenges for the design of cluster schedulers. A majority of resources are allocated to a few pretraining jobs, potentially causing head-of-line blocking issue and resulting in severe queuing delay. Existing DL cluster schedulers \cite{Gandiva,Pollux,Themis,Tiresias,ASTRAEA} typically depend on preemption mechanism, however, the considerable recovery overhead makes them not applicable to LLM workloads. This highlights the critical need for a scheduling system tailored for LLM clusters, considering the workload features of the entire pipeline.

\subsection{Workload Categories}
\label{sub_sec_workload_analysis}
To strive for a deeper understanding of the characteristics of different workloads in the LLM development pipeline (\S \ref{sub_sec_llm_pipline}), we further categorize jobs into specific types according to their production division and metadata (e.g., job names).

\noindent\textbf{Irrelevance of Job Count and Resource Usage}. Figure \ref{figure_treemap_job_types} presents the distribution of job counts and GPU time across various workload types, where only \Scluster contains SFT and MLLM workloads. Besides, MLLM jobs incorporate their own development pipeline (e.g., pretraining) and adopt smaller model scales for exploration purposes. Our analysis primarily focuses on LLM jobs.
It is obvious that evaluation jobs constitute the majority of the total job count in both clusters, yet they consume a relatively small portion of resources (0.8\% in \Kcluster). In contrast, pretraining jobs only account for 0.9\% and 3.2\% of the total job count but consume 69.5\% and 94.0\% of the total GPU time in \Scluster and \Kcluster respectively.

\noindent\textbf{Job Type Correlates with GPU Demand}. We further depict GPU demand distribution across various workload types in Figure \ref{figure_boxplot_job_types}. Each box is framed by the first and third quartiles, while the median value is indicated by the black line within the box. Both whiskers are defined at 1.5$\times$ the InterQuartile Range (IQR). Compared to evaluation jobs, which typically require less than 4 GPUs, pretraining jobs often require over 100 GPUs. This observation partially explains why evaluation jobs in \Kcluster consume only minimal resources in Figure \ref{figure_treemap_job_types}(d). Additionally, we notice that debugging jobs have a wide range of GPU requests, which aligns with the fact that testing jobs are typically needed for various types of tasks.

\noindent\textbf{Similar Temporal Distribution}.
Figure \ref{figure_cdf_job_types} shows the distribution of job duration and queuing delay across different workloads. In terms of job duration, although pretraining jobs have the longest duration, they surpass other workloads within an order of magnitude in the median, and less than 5\% jobs last for over 1 day in both clusters. This can be attributed to frequent failures during pretraining, which will be further explored in \S \ref{sec_failure}.
Regarding job queuing delay, contrary to previous reports \cite{Helios, MLaaS, Philly} suggesting that larger-scale jobs experience longer wait times, we observe that evaluation jobs have the longest queuing delay despite having the lowest GPU demands and shortest job duration. This discrepancy is due to the majority of resources being reserved for pretraining jobs to minimize their queuing delays. Evaluation jobs are typically submitted as a batch simultaneously with lower priority, utilizing the limited spare resources.

\begin{figure}[t]
    \vspace{-15pt}
    \centering
    \includegraphics[width=\linewidth]{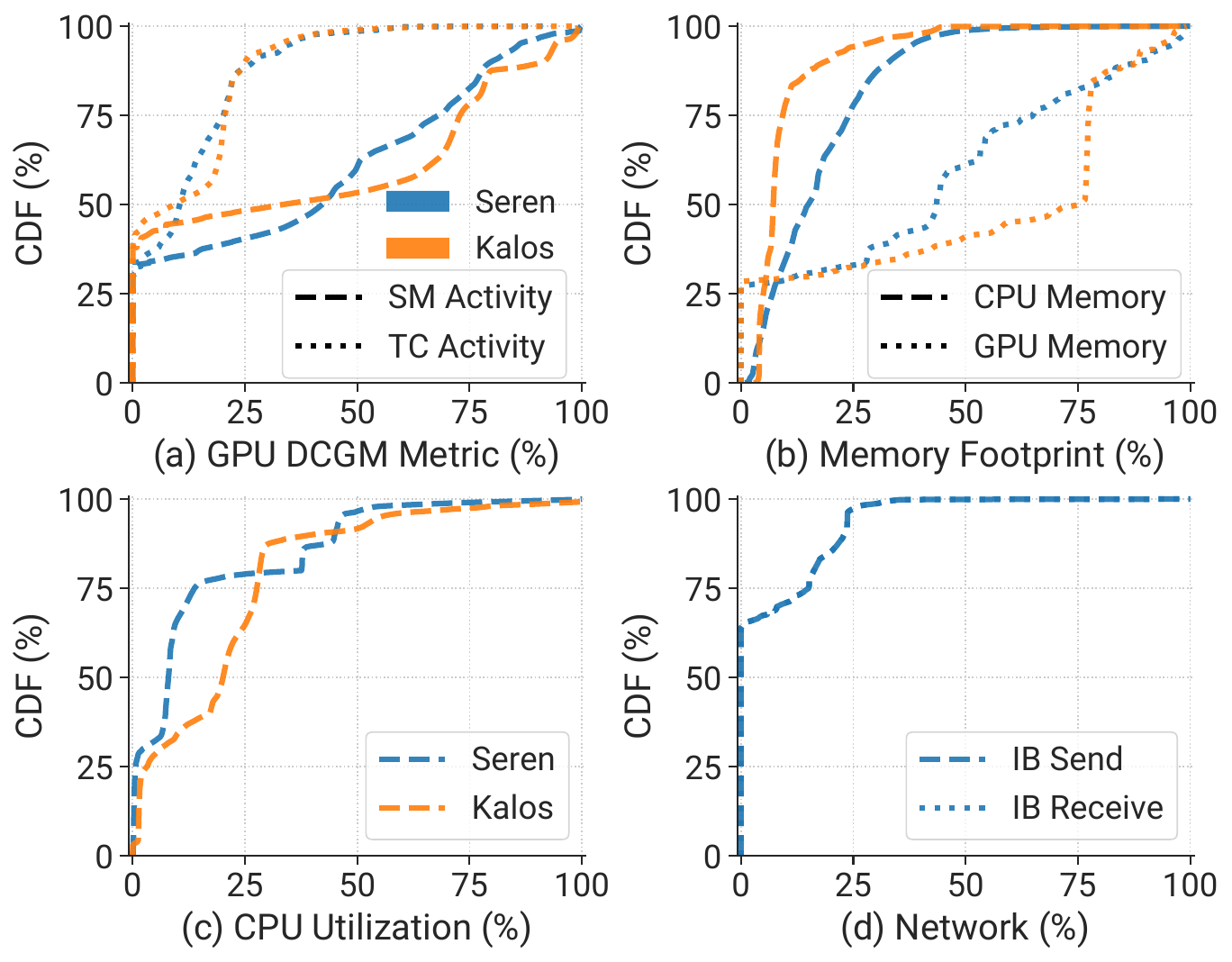}
    \caption{Infrastructure utilization. CDF of various metrics: (a) active fractions of Streaming Multiprocessor (SM) and Tensor Core (TC), (b) memory footprints of host and GPU, (c) CPU utilization, (d) normalized InfiniBand (IB) HCA send and receive bandwidths (\Scluster only). \Scluster and \Kcluster are represented by blue and orange lines respectively.}
    \label{figure_infra_util}
\end{figure}

\subsection{Infrastructure} 
\label{sub_sec_infrastructure_analysis}
Beyond the workload characterization, we further conduct a comprehensive analysis of our infrastructure utilization.

\noindent\textbf{Higher GPU Utilization}.
Given the critical role of GPUs in LLM development, as shown in Figure \ref{figure_infra_util} (a, b),  we collect fine-grained performance counter metrics from DCGM \cite{nvdcgm}, including SM Activity (\texttt{PROF\_SM\_ACTIVE}), TC Activity (\texttt{PROF\_PIPE\_TENSOR\_ACTIVE}), and GPU memory footprint (\texttt{DEV\_FB\_USED}). In contrast to PAI \cite{MLaaS}, where a significant portion of GPU memory is underutilized (less than 25\% memory), our observations in \Kcluster indicate that 50\% of GPUs consume over 75\% of GPU memory (60 GB). Furthermore, we observe that the median SM activity in both clusters is approximately 40\%, which is twice the reported 20\% in PAI. These findings align with the memory-intensive and compute-intensive natures of LLMs.

\begin{figure}[t]
    \vspace{-10pt}
    \centering
    \includegraphics[width=\linewidth]{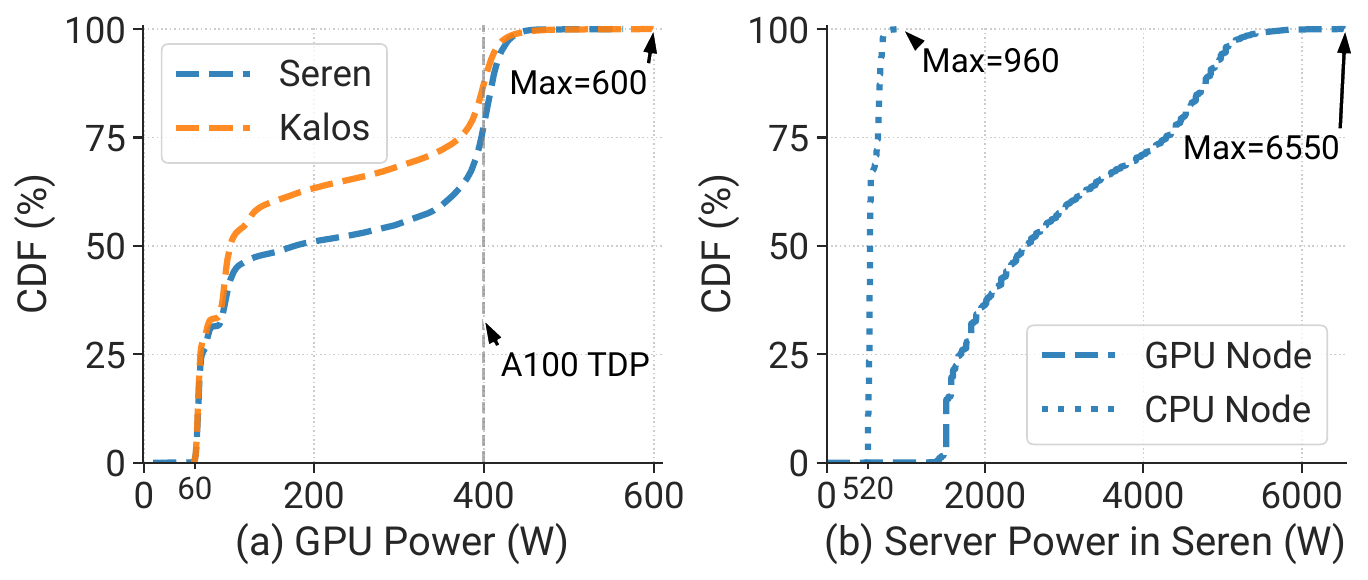}
    \caption{Power consumption. CDF of (a) A100 GPU power. (b) server power in \Scluster. \emph{TDP}: Thermal Design Power.}
    \label{figure_power_cdf}
\end{figure}

\begin{figure}[t]
    \vspace{-5pt}
    \centering
    \includegraphics[width=\linewidth]{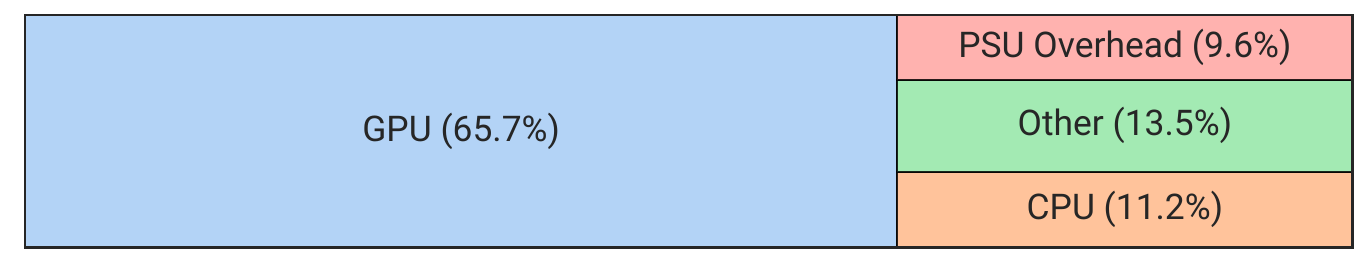}
    \caption{Average power distribution of hardware modules in \Scluster GPU servers, gathered from IPMI and DCGM.}
    \label{figure_power_treemap}
\end{figure}

\noindent\textbf{Underutilized Associated Resources}.
We also delve into the aspects of CPU, host memory, and network that are closely associated with LLM development. In Figure \ref{figure_infra_util} (b), we compare the memory footprint on the host side and GPU side. It is evident that CPU memory utilization remains below 50\%. Note that \Kcluster boasts twice the memory capacity (2TB) compared to \Scluster (Table \ref{table_cluster_summary}).
This demonstrates the significant underutilization of CPU memory. More detailed analysis is provided in Appendix \ref{appendix_subsec_cpu_mem}.
Although the GPU memory offloading technique \cite{ZeRO-Offload, ZeRO-infinity} improves CPU memory utilization and alleviates GPU memory limitations, it also impedes training throughput due to limited PCIe bandwidth. Therefore, we do not employ the offloading mechanism.
Additionally, due to a high CPU-to-GPU ratio (16 CPUs per GPU), CPUs are typically underutilized, as depicted in Figure \ref{figure_infra_util} (c). Moreover, in Figure \ref{figure_infra_util} (d), we measure the network send and receive bandwidths of IB in \Scluster. Two lines are well overlapped, as IB serves for symmetrical communication during LLM execution. We observe that NICs remain idle for over 60\% of the time, and the active bandwidth rarely surpasses 25\% of the maximum bandwidth provided by IB.

\subsection{Environmental Impact}
\label{sub_sec_environmental}
LLM development leads to substantial energy consumption and carbon emissions \cite{CarbonLLM, Zeus}. We report our analysis of infrastructure power consumption patterns to inspire future datacenter designs that minimize environmental impact.

\noindent\textbf{GPUs Dominate Power Consumption}.
Figure \ref{figure_power_cdf} (a) depicts the distribution of GPU power consumption. We observe that around 30\% of GPUs are in an idle state and still need to consume 60W. Besides, due to intensive computation demand, we find that 22.1\% and 12.5\% of GPUs consume over 400W (TDP) in \Scluster and \Kcluster respectively, with some even reaching 600W. This may cause the risk of some metastable issues \cite{Metastable}. Figure \ref{figure_power_cdf} (b) presents the power consumption distribution of all GPU servers, along with an additional 6 CPU-only servers, in \Scluster. We find GPU servers consume 5$\times$ power than CPU servers on average. Additionally, Figure \ref{figure_power_treemap} demonstrates that GPUs account for approximately 2/3 of the total power consumption in GPU servers, while CPUs only contribute 11.2\% and power supply units (PSUs) consume 9.6\% of the energy during voltage conversion. These observations align with the understanding that GPUs are the primary power consumers in LLM development. We also provide the estimation of datacenter carbon emission in Appendix \ref{appendix_subsec_carbon}.


%% file: 4_Profiling.tex
\section{Workload Profiling}
\label{sec_profile}
In this section, we conduct fine-grained analyses of resource utilization for representative tasks. Specifically, we focus on pretraining and evaluation jobs, as they are the most resource-intensive or quantity-intensive workloads.

\subsection{Pretraining Workload}
\label{sec_profile:pretrain}

\begin{figure}[t]
    \vspace{-10pt}
    \centering
    \includegraphics[width=\linewidth]{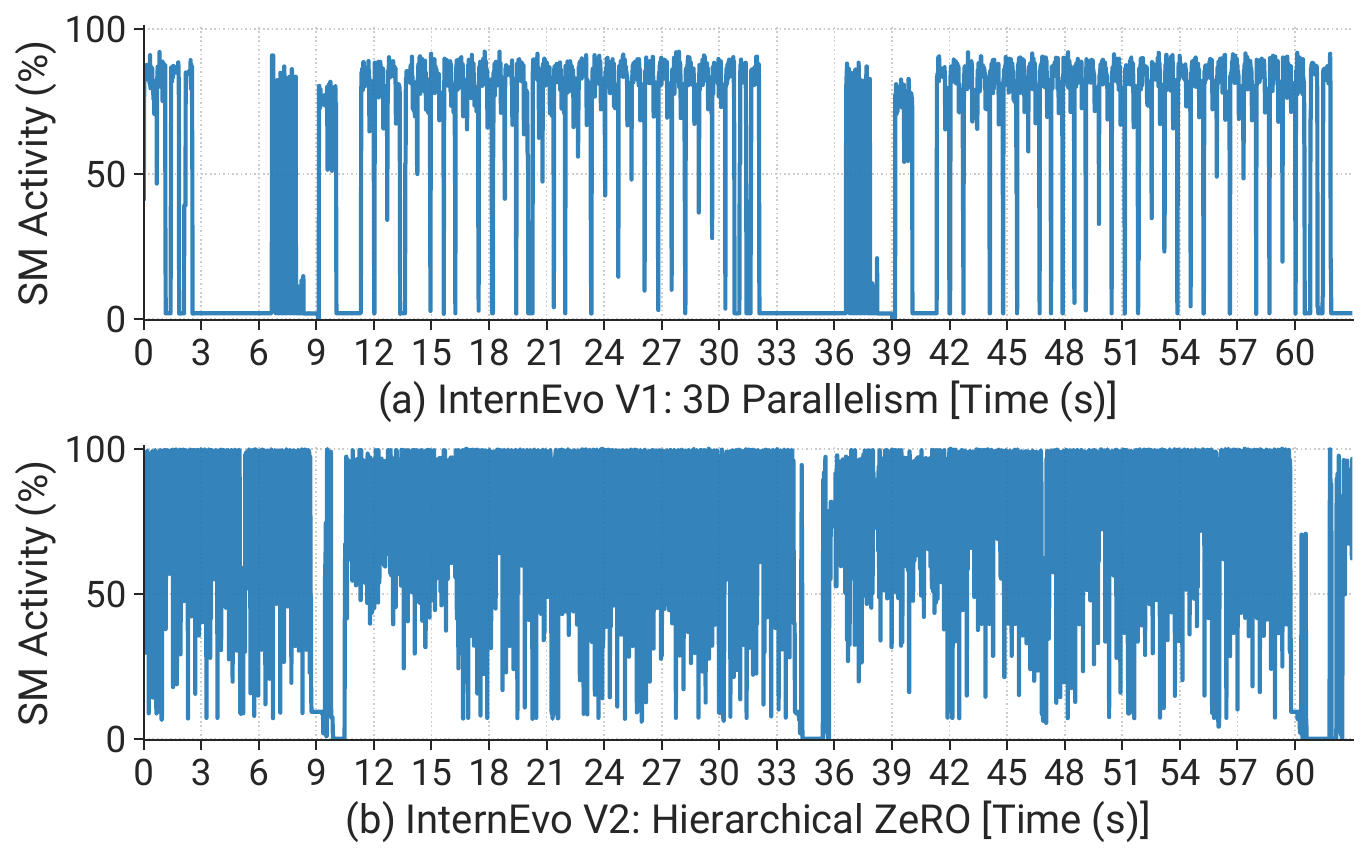}
    \caption{GPU SM utilization of pretraining a 123B LLM using different strategies of InternEvo \cite{InternEvo} over 2048 GPUs.}
    \label{figure_profile_pretrain}
\end{figure}

As aforementioned, pretraining LLMs requires substantial computational resources. To enhance training efficiency, our pretraining framework, InternEvo \cite{InternEvo}, undergoes continuous refinement and iteration in its system design. As presented in Figure \ref{figure_profile_pretrain}, the initial version of InternEvo (adopted by our early jobs) is denoted as (a) primarily utilizes 3D parallelism akin to that of MegatronLM \cite{MegatronLM}, and (b) employs a hierarchical ZeRO mechanism \cite{InternEvo} that implements selective redundant sharding of model states. To provide a detailed example, we profile an LLM with 123 billion parameters across 2048 GPUs. We also provide the profiling results of 1024 GPUs in Appendix \ref{appendix_subsec_pertraining}. For (a) 3D parallelism approach, we adopt a configuration with \texttt{pipeline parallelism}$=4$, \texttt{tensor parallelism}$=8$. We sample the first GPU of the first pipeline rank for profiling.
For (b) hierarchical ZeRO approach, we limit parameter sharding to subgroups of 64 GPUs each and enable recomputation.
We collect GPU performance counters like DCGM metrics at 1 ms intervals.

\begin{figure}[t]
    \vspace{-10pt}
    \centering
    \includegraphics[width=\linewidth]{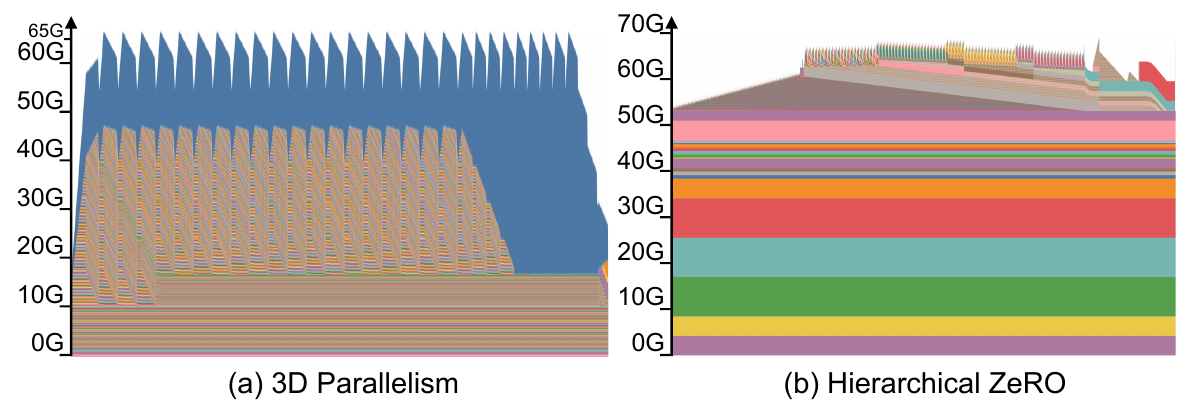}
    \caption{Memory snapshot under different pretraining strategies. Note that the extensive blue segment at the top of (a) is simplified and can be further broken down into massive fragments (memory allocations), similar to the lower part.}
    \label{figure_mem_timeline}
\end{figure}

\begin{figure}[t]
    \vspace{-5pt}
    \centering
    \includegraphics[width=0.7\linewidth]{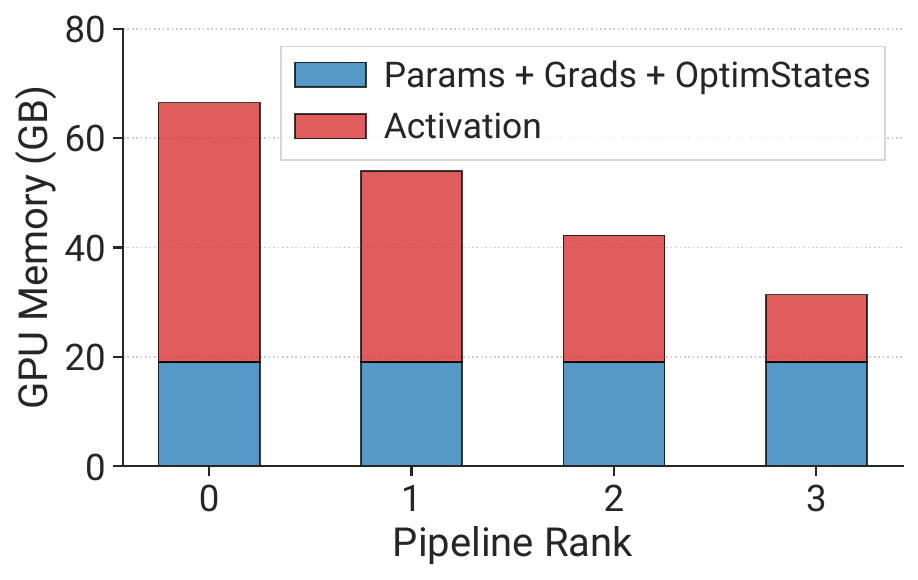}
    \caption{GPU memory consumption of different pipeline ranks employing the 1F1B \cite{PipeDream}  strategy in InternEvo V1.}
    \label{figure_activation_pp}
\end{figure}

\noindent\textbf{GPU SM Utilization}.
Figure \ref{figure_profile_pretrain} illustrates the GPU SM  utilization for the same LLM under various training strategies. Both versions maintain the same global batch size and are optimized according to their respective configurations. It is evident that InternEvo V2 presents superior peak SM utilization and exhibits reduced idle periods compared to InternEvo V1, achieving around 16\% acceleration. The relatively low utilization of 3D parallelism is mainly due to the impact of communication introduced by hybrid parallelism on the critical path, such as bubbles in pipeline parallelism. Note that the different inter- and intra-node communication hardware settings may lead to different optimal configurations.


\noindent\textbf{GPU Memory Footprint}.
For a model comprising $\Psi$ parameters, in the mainstream mixed precision training using Adam \cite{Adam} optimizer, the memory footprint of the parameters, gradients, and optimizer states are 2$\Psi$, 2$\Psi$, and 12$\Psi$, respectively. To reduce memory cost, ZeRO \cite{ZeRO} effectively shards redundant memory of these elements across global GPU workers.
Figure \ref{figure_mem_timeline} illustrates the actual GPU memory usage over time captured by the Pytorch memory snapshot tool \cite{PytorchMemShot}. The upper dynamic part represents activations and gradients, while the lower static part represents the memory occupied by parameters and optimizer states. Note that only allocated memory is depicted, while reserved memory is not presented. Our analysis reveals that, in comparison to hierarchical ZeRO, the memory requirement for activations in 3D parallelism is substantially higher. This observation underscores the importance of efficient activation memory management as a key factor for enhancing batch size and throughput in 3D parallelism.

\begin{figure}[t]
    \vspace{-10pt}
    \centering
    \includegraphics[width=\linewidth]{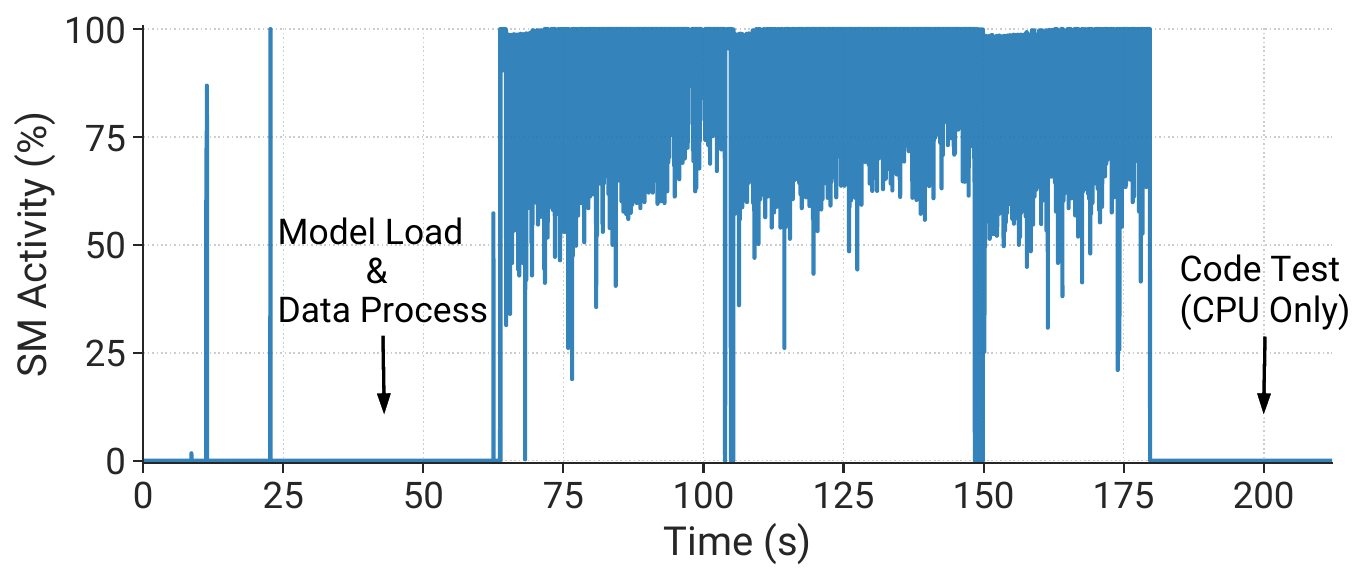}
    \caption{GPU SM utilization for the entire evaluation workload on HumanEval \cite{HumanEval} dataset using a 7B LLM.}
    \label{figure_profile_eval}
\end{figure}

\noindent\textbf{Imbalance in Activation Sizes}.
When employing pipeline parallelism, each rank needs to hold a different quantity of activations since the diverse number of micro-batches pending backward computation across various pipeline ranks.
Figure \ref{figure_activation_pp} illustrates this imbalance issue on different pipeline ranks. It suggests that we should employ a specialized partitioning mechanism to address the unbalanced memory usage among different ranks in pipeline parallelism, in order to achieve higher efficiency, such as recomputing activations.


\subsection{Evaluation Workload}
\label{sec_profile:evaluation}

It is necessary to regularly evaluate the checkpoints produced during pretraining to guide the evolution of LLM pretraining. Therefore, the LLM evaluation jobs take the majority of jobs, each performing metric computation on different LLM benchmark datasets. We analyze the workflow of the entire evaluation and combine it with fine-grained resource usage information collection quantitatively, demonstrating two upcoming resource utilization issues. We will also discuss the corresponding solutions in \S \ref{sub_sec_system_eval}.

\noindent\textbf{High Model Loading and Data Preprocessing Overhead}. During the initiation phase of evaluation jobs, it is imperative to load model checkpoints for each task. Additionally, the data preprocessing stage, particularly for tokenization, constitutes a significant time expenditure. These factors contribute to the underutilization of allocated GPU resources for a relatively long period. As illustrated in Figure \ref{figure_profile_eval},  the evaluation task consumes over 1 minute prior to the actual GPU inference, accounting for 29.5\% of the evaluation duration. This overhead is likely to increase with larger models or datasets. To address the preprocessing overhead, one effective strategy is to cache the tokenized data. Moreover, evaluation jobs are flexible, allowing for the consolidation of multiple evaluation tasks (datasets) into a single job. This consolidation can effectively reduce the relative time overhead of the model loading phase within the evaluation process.

\noindent\textbf{High Metric Computation Overhead}.
The evaluation process can often involve complex and time-consuming metric computation. For example, synthesized program correctness tests need to be performed on coding datasets like HumanEval \cite{HumanEval} and MBPP \cite{MBPP}. Moreover, the OpenAI GPT-4 API is invoked to assess the performance of model conversations (e.g., Chatbot Arena \cite{Arena}). These procedures can take up to 30 minutes, during which the GPU remains idle.
Therefore, we can observe distinct stages of GPU usage, including stages that require GPU for inference and generation, and stages that do not require GPU for metric computation and verification.
Taking the HumanEval benchmark as an example, as shown in Figure~\ref{figure_profile_eval}, the GPU is idle for the last 42 seconds, wasting about 19.0\% of the total GPU time.

\begin{table*}[t]
    \centering
    \vspace{-15pt}
    \renewcommand{\arraystretch}{1.1}
    \resizebox{\linewidth}{!}{
        \begin{tabular}{clrrrrrrrrrrc}
            \toprule
            \multirow{2}{*}{\textbf{Category}} & \multicolumn{1}{c}{\multirow{2}{*}{\textbf{Reason}}} & \multicolumn{1}{c}{\multirow{2}{*}{\textbf{Num}}} & \multicolumn{2}{c}{\textbf{GPU Demand}} & \multicolumn{2}{c}{\textbf{Time to Failure (mins)}} & \multicolumn{2}{c}{\textbf{GPU Time (mins)}} & \multicolumn{3}{c}{\textbf{Time to Restart (mins)}} & \multirow{2}{*}{\textbf{Cluster}}                                                 \\ \cmidrule(lr){4-5}  \cmidrule(lr){6-7} \cmidrule(lr){8-9} \cmidrule(lr){10-12}
                                               & \multicolumn{1}{c}{}                                 & \multicolumn{1}{c}{}                              & Average                                 & Median                                              & Average                                      & Median                                              & Average                           & Total\% & Average & Median & TR/TF\%   &      \\ \midrule
            \multirow{9}{*}{Infrastructure}    & NVLink Error                                         & 54                                                & 800                                     & 896                                                 & 868.1                                        & 155.3                                               & 585683                            & 30.25\% & 95.6    & 0.2    & 11.02\%   & S, K \\
                                               & CUDA Error                                           & 21                                                & 847                                     & 1024                                                & 923.2                                        & 586.0                                               & 785099                            & 15.77\% & 78.3    & 2.0    & 8.48\%    & S, K \\
                                               & Node Failure                                         & 16                                                & 712                                     & 768                                                 & 1288.8                                       & 535.8                                               & 934394                            & 14.30\% & 102.8   & 21.5   & 7.98\%    & S    \\
                                               & ECC Error                                            & 12                                                & 680                                     & 512                                                 & 1303.4                                       & 1192.3                                              & 958404                            & 11.00\% & 2.8     & 1.8    & 0.21\%    & S, K \\
                                               & Network Error                                        & 12                                                & 758                                     & 768                                                 & 549.6                                        & 310.1                                               & 394821                            & 4.53\%  & 592.1   & 7.4    & 107.74\%  & S, K \\

                                               & Connection Error                                     & 147                                               & 29                                      & 1                                                   & 51.9                                         & 0.5                                                 & 24492                             & 3.44\%  & 0.8     & 0.0    & 1.51\%    & S, K \\
                                               & S3 Storage Error                                     & 10                                                & 422                                     & 256                                                 & 2317.8                                       & 202.2                                               & 222151                            & 2.12\%  & 6.2     & 0.2    & 0.27\%    & S    \\
                                               & NCCL Timeout Error                                   & 6                                                 & 596                                     & 512                                                 & 159.7                                        & 48.1                                                & 86856                             & 0.50\%  & 66.7    & 43.6   & 41.78\%   & K    \\
                                               & NCCL Remote Error                                    & 3                                                 & 1152                                    & 1024                                                & 50.5                                         & 22.6                                                & 52419                             & 0.15\%  & 0.0     & 0.7    & 0.09\%    & K    \\ \hline
            \multirow{9}{*}{Framework}         & Dataloader Killed                                    & 6                                                 & 445                                     & 508                                                 & 1580.6                                       & 961.4                                               & 764170                            & 4.38\%  & 115.1   & 0.9    & 7.28\%    & K    \\
                                               & Attribute Error                                      & 67                                                & 228                                     & 8                                                   & 67.8                                         & 1.2                                                 & 60914                             & 3.90\%  & 2.4     & 0.0    & 3.58\%    & S, K \\
                                               & Out of Memory Error                                  & 14                                                & 572                                     & 640                                                 & 323.8                                        & 14.5                                                & 245278                            & 3.28\%  & 122.7   & 1.2    & 37.89\%   & S, K \\
                                               & Runtime Error                                        & 65                                                & 441                                     & 352                                                 & 66.4                                         & 3.9                                                 & 27667                             & 1.72\%  & 10.9    & 1.5    & 16.41\%   & S, K \\
                                               & Assertion Error                                      & 105                                               & 413                                     & 256                                                 & 41.7                                         & 3.0                                                 & 12315                             & 1.24\%  & 185.9   & 1.6    & 445.87\%  & S, K \\
                                               & Value Error                                          & 33                                                & 387                                     & 256                                                 & 9.9                                          & 3.7                                                 & 5049                              & 0.16\%  & 27.4    & 0.6    & 276.74\%  & S, K \\
                                               & Zero Division Error                                  & 5                                                 & 499                                     & 256                                                 & 14.5                                         & 15.6                                                & 5363                              & 0.03\%  & 2.5     & 1.1    & 17.31\%   & S, K \\
                                               & Model Loading Error                                  & 104                                               & 8                                       & 8                                                   & 2.6                                          & 2.6                                                 & 20                                & 0.00\%  & 0.0     & 0.0    & 0.00\%    & K    \\
                                               & Dataset Loading Error                                & 5                                                 & 1                                       & 1                                                   & 1.6                                          & 1.6                                                 & 1                                 & 0.00\%  & 0.0     & 0.0    & 0.00\%    & K    \\ \hline
            \multirow{11}{*}{Script}           & File Not Found Error                                 & 568                                               & 21                                      & 1                                                   & 14.2                                         & 0.4                                                 & 5210                              & 2.83\%  & 0.4     & 0.0    & 2.58\%    & S, K \\
                                               & OS Error                                             & 266                                               & 8                                       & 1                                                   & 9.6                                          & 0.8                                                 & 1098                              & 0.28\%  & 0.3     & 0.0    & 3.17\%    & S, K \\
                                               & Type Error                                           & 620                                               & 18                                      & 4                                                   & 0.9                                          & 0.3                                                 & 97                                & 0.06\%  & 0.2     & 0.0    & 28.27\%   & S, K \\
                                               & Name Error                                           & 18                                                & 247                                     & 24                                                  & 3.2                                          & 0.5                                                 & 947                               & 0.02\%  & 2.9     & 2.4    & 90.92\%   & S, K \\
                                               & Permission Error                                     & 7                                                 & 438                                     & 512                                                 & 4.3                                          & 0.8                                                 & 2131                              & 0.01\%  & 2.4     & 2.2    & 56.38\%   & S    \\
                                               & Import Error                                         & 111                                               & 93                                      & 8                                                   & 1.1                                          & 0.4                                                 & 74                                & 0.01\%  & 0.7     & 0.0    & 63.68\%   & S, K \\
                                               & Key Error                                            & 260                                               & 7                                       & 0                                                   & 3.0                                          & 1.6                                                 & 55                                & 0.01\%  & 0.1     & 0.0    & 2.10\%    & S, K \\
                                               & Syntax Error                                         & 10                                                & 391                                     & 384                                                 & 0.7                                          & 0.6                                                 & 348                               & 0.00\%  & 1.7     & 1.7    & 261.73\%  & S, K \\
                                               & Argument Error                                       & 3                                                 & 344                                     & 512                                                 & 0.7                                          & 0.7                                                 & 288                               & 0.00\%  & 2.7     & 0.7    & 408.47\%  & S    \\
                                               & Called Process Error                                 & 4                                                 & 256                                     & 256                                                 & 0.2                                          & 0.2                                                 & 52                                & 0.00\%  & 11.7    & 10.9   & 5714.29\% & S    \\
                                               & Index Error                                          & 23                                                & 6                                       & 1                                                   & 1.6                                          & 0.9                                                 & 21                                & 0.00\%  & 0.8     & 0.0    & 49.73\%   & S, K \\ \hline
        \end{tabular}}
    \caption{Job failure statistics. It is sorted based on \textbf{Total\%} (i.e., the percentage of GPU time summation in different categories). \textbf{Num}: Number of Occurrence. \textbf{TF}: Time to Failure.  \textbf{TR}: Time to Restart (i.e., Restart Timestamp $-$ Failure Timestamp). \textbf{GPU Time}: TF$\times$GPU Demand. \textbf{S}/\textbf{K}: Occurrence of errors in \Scluster/\Kcluster respectively.}
    \label{table_failure_summary}
    \vspace{-10pt}
\end{table*}

%% file: 5_Failure.tex
\section{Failure Analysis}
\label{sec_failure}

In this section, we conduct a comprehensive analysis of job failures, primarily relying on \emph{runtime logs} and \emph{hardware monitor data} from our two clusters. In \Kcluster, we gather logs from 32,500 tasks, which include 31,293 (96.3\%) inference tasks, 647 (2.0\%) pretraining tasks, and debugging tasks (1.7\%). In \Scluster, we only collected logs from 675 pretraining tasks. Additionally, for pretraining tasks, we extract all pertinent information and metadata recorded in the logs, including actual training steps, cold-start overhead, recovery timestamp, etc. We hope our analysis can provide insights for future fault-tolerance research in the development of LLMs.





\subsection{Failure Category}

We employ a failure diagnosis system, leveraging a combination of rule-based and LLM techniques, to extract error information from the runtime logs. We provide detailed explanations of this system in \S \ref{sub_sec_system_pretrain}. Besides, to ensure the accurate identification of the types and root causes of failures, manual checks and corrections are conducted. Table \ref{table_failure_summary} provides a summary of common failures in \DC, including their occurrence frequency and restart time. Basically, they can be classified into three categories as follows. Note that these classifications may overlap, and the primary criterion for classifying a specific type of error is its most frequent occurrence.

\begin{itemize}[leftmargin=*,topsep=0pt, itemsep=-3pt, itemindent=8pt]
    \item \textbf{Infrastructure.} Infrastructure-related failures arise from issues within the underlying computational platform or remote storage. These failures mainly occur midway through the job execution process, especially in pretraining tasks. They severely impact the training progress due to laborious and time-consuming recovery process.

    \item \textbf{Framework.} Several types of runtime errors, such as \textsl{RuntimeError}, \textsl{ValueError}, and \textsl{AttributeError}, can be associated with tensor operations, shapes, data types, or unexpected behaviors. They are often observed in the initial phases of jobs and are typically resolved by fixing the configurations.

    \item \textbf{Script.} Script errors typically stem from programming errors or user oversights. They constitute the majority of failures and are often addressed by revising codes.
\end{itemize}

\begin{figure}[t]
    \centering
    \vspace{-10pt}
    \includegraphics[width=\linewidth]{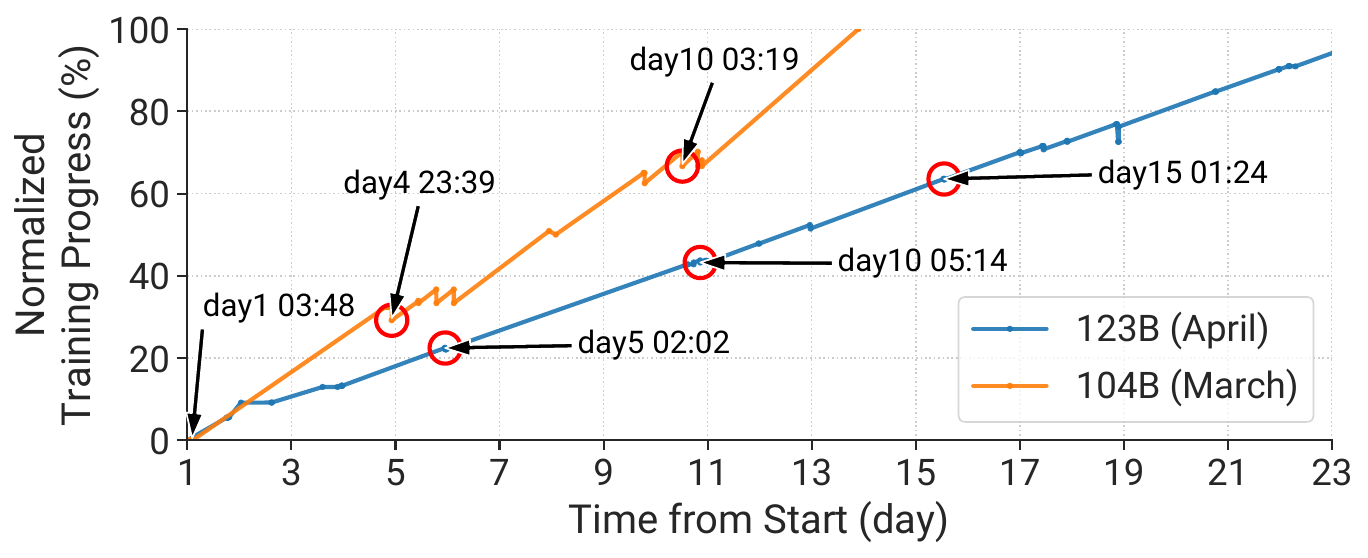}
    \caption{The training progress of two LLMs, with special emphasis on the manual recovery at night.}
    \label{figure_failure_example}

\end{figure}

\subsection{Failure Characterization}

We highlight several key observations from our failure analysis. Besides, we share our
experience in troubleshooting and resolving some intriguing framework problems in \S \ref{appendix_lessons}.

\noindent\textbf{Infrastructure Failures Cause Most Severe Impact}.
As shown in Table \ref{table_failure_summary}, jobs that fail because of infrastructure issues often use a substantial number of GPUs (\textit{GPU Demand}) and require considerable effort to restart (\textit{Time to Restart}). They take over 82\% GPU resources (\textit{GPU Time}) with only 11\% failed job quantity (\textit{Num}). Most of these jobs are long-term pretraining tasks that can experience hardware failures multiple times, such as issues with GPU (e.g., \textsl{CUDAError}, \textsl{ECCError}), NVLink (\textsl{NVLinkError}), and network system (\textsl{NCCLRemoteError}, \textsl{S3StorageError}). Note that \textsl{NodeFailure} indicates uncategorized errors caused by unclear hardware issues. Addressing these infrastructure failures requires meticulous diagnostic efforts to pinpoint the source of the problems, often leading to the maintenance or replacement of defective hardware, which results in significant restart costs.

\noindent\textbf{Failures Caused by High Temperature}. Another noteworthy observation is that training 7B models in \Kcluster tend to result in GPU overheating, which can cause \textsl{NVLinkError} or \textsl{ECCError}. This phenomenon is largely due to the highly optimized communication cost, resulting in an exceptionally low GPU idle rate. We observe that the overall temperature in the cluster server room increased by approximately $5^\circ$C when training these models. Besides, we find most of these jobs occurred in July 2023, which is the hottest month on record \cite{HottestMonth}. This anomalous climate may be a potential cause of these failures, which is aligned with the finding recently reported by Microsoft \cite{Anubis}. We provide more detailed data on GPU temperature in Appendix \ref{appendix_subsec_temperature}.
Subsequently, our team enhanced the cooling capabilities of the cluster, leading to a significant reduction in the frequency of such failures.

\noindent\textbf{Many Failures Induced by Auxiliary Services}. In our pretraining framework, we connect to external components or services for metric reporting, logging, monitoring and alerting. These auxiliary services are vulnerable to network instabilities, potentially resulting in timeouts or failures that can decelerate or disrupt the training process. A significant number of \textit{ConnectionError} and \textit{NetworkError} incidents stem from these auxiliary services.

\noindent\textbf{Evaluation Jobs Rarely Encounter Errors}.  In \Kcluster, only 6.7\% of evaluation tasks encounter errors, and notably, there are no recorded instances of GPU or NVLink failures. The low error rate may be attributed to their short duration and the resultant decreased stress on GPUs and NVLink connections. Consequently, this diminishes the chance of hardware and operational failures that are more frequent in pretraining jobs.


\subsection{Failure Recovery}
\label{failure_recovery}
There are three scenarios where we should restart a job: (1) when an error occurs within the job, (2) when there are anomalies in training metrics such as a loss spike, and (3) when the training process is stuck. A `loss spike' refers to a sudden increase in the loss that was previously decreasing normally, and does not recover over a certain period. Upon restarting, jobs revert to the last checkpoint, resulting in loss of training progress. Since existing LLM frameworks lack automatic recovery support, developers usually manually restart interrupted training jobs. Developers often need to be on call in turn to ensure timely completion of the pretraining model.

As shown in Figure \ref{figure_failure_example}, we select two pretraining jobs in the early stage (March to April) when we handle all failures manually. We extract information from the logs of two clusters' large-scale model training processes, including the runtime duration of each submission, start and end times, and the initial and final iteration numbers of training.
The 104B model is an early attempt when the framework is still under development. Consequently, the process of loading previous model checkpoints led to a substantial loss in the overall training process.
Conversely, in the training of the 123B model a month later, we improved the framework and adopted smaller checkpoint save intervals. Moreover, we added a feature to gracefully terminate jobs, allowing for the preservation of current training results before ending the job. It is evident that the training process of the 123B model is more stable, with fewer losses incurred due to rollbacks. However, this progress came at a cost, as jobs that were interrupted at various times had to be rapidly restarted.

%% file: 6_System.tex
\section{Deployed LLM Systems}
\label{sec_implication}

As highlighted earlier, the development process of LLMs presents significant obstacles yet unveils viable strategies for overcoming these issues. This section will introduce our efforts in two stages: (1) \emph{Pretraining}: enhancing fault tolerance through LLM-involved failure diagnosis and automatic recovery. (2) \emph{Evaluation}: achieving prompt performance response via task decomposition.

\subsection{Fault-tolerant Pretraining}
\label{sub_sec_system_pretrain}
\noindent\textbf{Motivation}. During LLM pretraining, failures are inevitable and frequently occur due to the substantial number of GPUs involved and the extensive duration of the training process \cite{Bamboo, GEMINI, Oobleck, Varuna}. These failures can dramatically impede
the training progress and lead to severe resource inefficiency (\S \ref{sec_failure}). Consequently, to minimize infrastructure downtime, it is common practice to assign on-call duties to address failures manually. This places a significant burden on engineers and researchers, as expressed in the complaints raised by the Meta OPT \cite{OPT} and BigScience BLOOM \cite{BLOOM} teams. Our team also faces this problem. To alleviate this burden and enhance hardware efficiency, we develop a system that automatically detects the root causes of faults and facilitates recovery.

\noindent\textbf{System Design}.
Our fault tolerance system is seamlessly integrated into our LLM pretraining framework. It comprises three essential modules: (1) \emph{Checkpointing}, achieving more frequent model saving to minimize the loss of training progress; (2) \emph{Diagnosis}, using a combination of heuristic rules and LLM to identify the root cause of different failures accurately; (3) \emph{Recovery}, employing a holistic detection toolkit to pinpoint fault nodes and automatically restart training from the properly saved checkpoint. We delve into them in detail.

\noindent\textbf{\textsl{1. Asynchronous Checkpointing}}.
Frequent checkpointing efficiently mitigates the wasted time caused by unexpected faults \cite{Check-N-Run}. However, as LLMs can produce TB-scale model states (referring to total model states across all GPUs), the process of saving checkpoints itself can introduce substantial overhead, resulting in training time slowdown up to 43\% \cite{CPR}. To tackle this problem, we adopt the asynchronous checkpointing strategy \cite{CheckFreq, DeepFreeze}, which effectively separates the checkpointing process from the training process. Our observations indicate that the CPU memory (refer to Figure \ref{figure_infra_util} (b)) is capable of accommodating several checkpoints. By taking advantage of this, we can store the model state in memory and utilize a separate thread to regularly save these states to remote persistent storage. This simple strategy significantly reduces checkpointing overhead.

\begin{figure}[t]
    \centering
    \includegraphics[width=\linewidth]{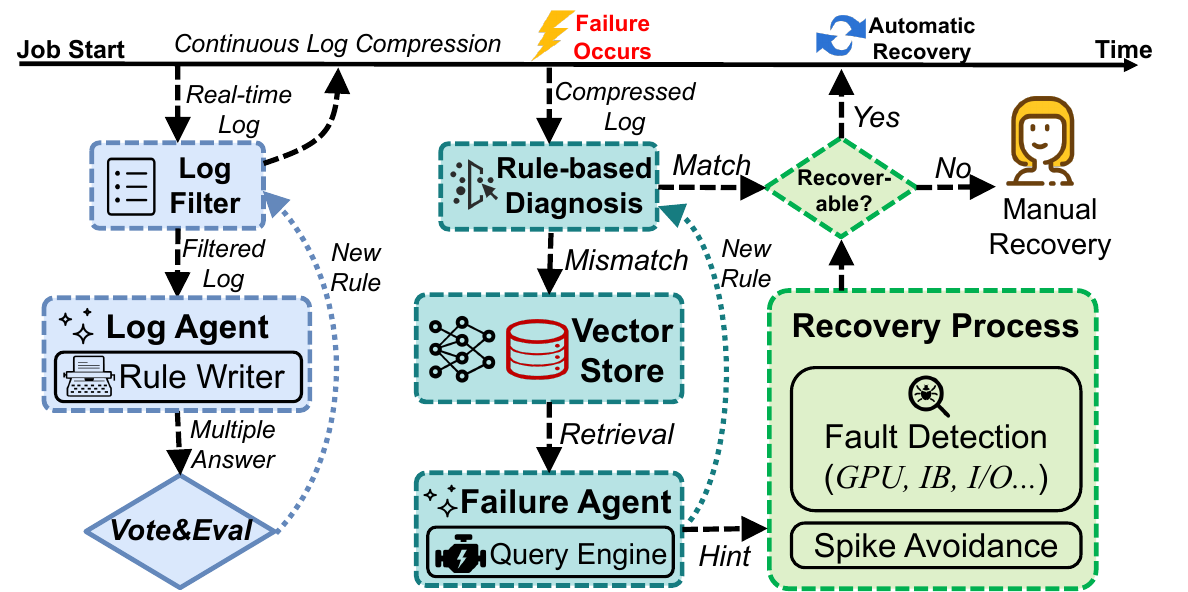}
    \caption{Workflow of failure diagnosis and model recovery.}
    \label{figure_system_pretrain}
\end{figure}

\noindent\textbf{\textsl{2. Failure Diagnosis}}.
As we discussed in \S \ref{sec_failure}, failures can arise from numerous intricate factors, including errors from user script or framework, as well as issues with hardware subjected to high-stress conditions. It is crucial to determine whether a failure is recoverable for the purpose of automatic recovery. A common approach is to use a combination of heuristic rules to filter and conduct regular expression matching on the logs of faulty jobs \cite{Philly, DISTALYZER, LogFailures, LogParsing, LogPPT}. However, this approach often proves inaccurate due to the wide-ranging diversity and complexity of error logs. There might not be a specific error statement in many cases, but multiple errors could coexist simultaneously. For example, a job might fail with messages that include \textsl{NCCLTimeoutError}, \textsl{CUDAError}, and multiple kinds of \textsl{RuntimeError}, whereas the root cause is \textsl{CUDAError}. Trying to match every error scenario with a specific rule set can become impractical.

To address this challenge, we utilize the exceptional text understanding ability and extensive knowledge base of LLMs to identify the root causes of different failures automatically. As depicted in Figure \ref{figure_system_pretrain}, we incorporate an LLM with rule-based diagnosis to achieve efficient and accurate failure diagnosis.
It mainly contains the following two steps:

\ding{228}\textbf{Real-time Log Compression}.  The extensive log files generated by pretraining jobs, primarily consisting of training metric records, can reach sizes of hundreds of MBs. To accelerate diagnosing and meet the context length limit of LLMs, log compression is conducted first. The system continuously updates a collection of regular expressions, termed as \textit{Filter Rules}.
These rules efficiently remove regular log outputs, such as initialization information, training metric records, framework outputs, and debug information. A vital component of the system, the LLM-based \textit{Log Agent}, is responsible for analyzing real-time generated log segments and identifying lines that follow fixed patterns. By doing so, the LLM-based \textit{Log Agent} dynamically writes regular expressions to update the \textit{Filter Rules}, effectively minimizing the size of the log files. Additionally, the \textit{Log Agent} forwards identified error messages to subsequent modules for diagnosis.

Furthermore, we employ the self-consistency \cite{SelfConsistency} approach to ensure the robustness of the \textit{Log Agent}'s results and to guarantee the formatting of these results. This involves processing each log segment multiple times and having another LLM vote on multiple results from the \textit{Log Agent}, ensuring the accuracy of matches through regular expressions. Over time, the \textit{Filter Rules} become more comprehensive for the current task, making the log filtering process more efficient. Furthermore, the system can utilize metadata from tasks to identify repetitive or similar tasks, directly applying existing \textit{Filter Rules} for log filtering, thereby avoiding redundant work. This feature is particularly beneficial in large model cluster environments, where fewer tenants and task resubmissions are common.

\ding{228}\textbf{LLM-assisted Automated Diagnosis}. The \textit{Log Agent} efficiently compresses runtime logs, isolating critical error logs like \textsl{CUDAError}s or runtime exceptions.
Though logs are already compressed upon arrival at this module, error logs may still be lengthy. We apply a two-step approach to tackle this issue. First, the error logs are compared against a rule set that has been defined over time through the diagnosis of errors from past failed jobs. If the pre-defined rules fail to diagnose the issue, the compressed log is vectorized through an embedding model and stored in a vector store, serving as a retrieval repository. Then, the \textit{Failure Agent} intervenes. It utilizes a \textit{Query Engine} \cite{RAG} to search through the vector store. Through this search, the \textit{Failure Agent} can identify log lines that reflect the root cause of job interruption, extract the type of error, and indicate whether the error originated from user mistakes or infrastructure failures, providing a hint for the recovery process. In addition, it also generates a mitigation suggestion for users or the operation team. 

The \textit{Failure Agent} also contributes to the continuous learning of the failure diagnosis system. For each new failure, once diagnosed and resolved, the \textit{Failure Agent} writes a corresponding regular expression and adds it to the \textit{Rule-based Diagnosis} module. This process is iterative and ensures that the \textit{Failure Diagnosis} system evolves, becoming more adept at diagnosing and suggesting mitigation methods for failures.
To achieve more robust performance, we currently utilize the GPT-4 \cite{ChatGPT} for diagnosis, with plans to transition to our LLMs.

\noindent\textbf{\textsl{3. Fast Fault Detection and Recovery}}.
Based on the failure diagnosis result, if it belongs to one kind of infrastructure failure, we conduct a corresponding detection test to identify the problematic nodes. For instance, to promptly resolve the frequent \textsl{NVLinkError}, we employ a two-round  NCCL test approach akin to that utilized by DLRover \cite{DLRover}. First, we divide all nodes into multiple two-node worlds and execute \texttt{allgather} task in each pair. If the total number of servers is odd, we leave one world size as three.  If \texttt{allgather} task fails in a world, the nodes in that world are potentially faulty nodes. Then, in the second round, we pair potential faulty nodes with normal nodes to form new worlds. The nodes in each world continue to execute the \texttt{allgather} task, thus identifying the faulty nodes and then cordoning them off. On the other hand, if the failure is attributed to a sudden increase in loss (i.e.,  `loss spike' \cite{OPT, PaLM}), which is automatically triggered by our pretraining framework, we opt to an earlier healthy restart checkpoint and bypass subsequent data batches. This approach effectively maintains model quality.


\noindent\textbf{System Performance}.
Our asynchronous checkpointing strategy offers a substantial reduction in checkpointing overheads, as the checkpointing process does not block the training process. The checkpoint time and overhead percentage of 7B and 123B size models are reduced by 3.6$\sim$58.7$\times$ (interval=30 mins), respectively. Note that the time taken for persisting to storage is not included in asynchronous checkpointing measurement. Moreover, our failure diagnosis system significantly reduces manual intervention by around 90\% and thus reduces developers' burden. Note this is not a rigorous assessment since components of our system are still under improvement.


\begin{figure}[t]
    \centering
    \includegraphics[width=\linewidth]{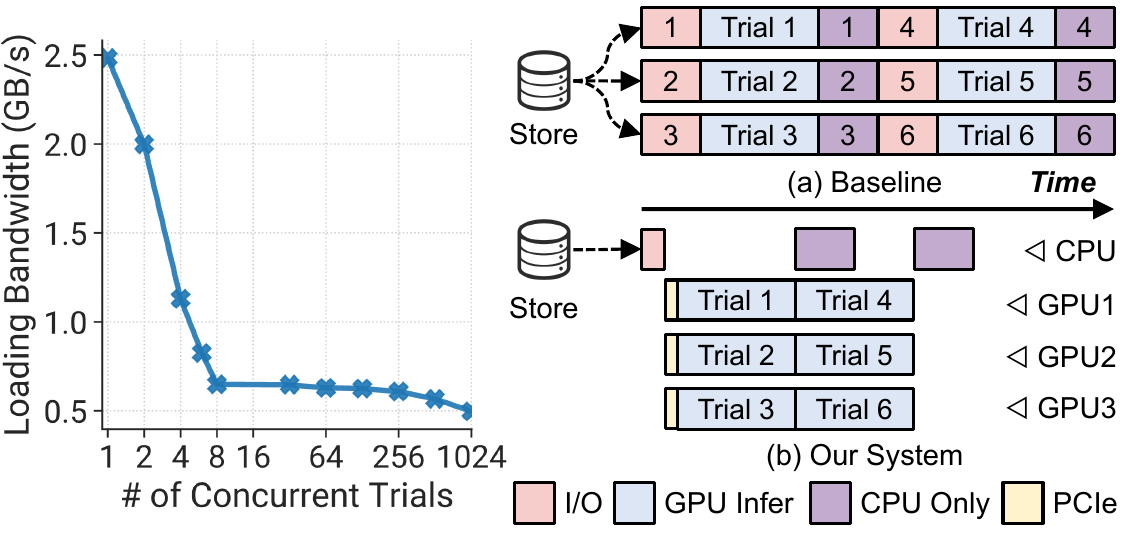}
    \caption{\emph{Left}: Stress testing of model loading from remote storage in \Scluster. Each trial involves one GPU. \emph{Right}: Scheduling evaluation trials. (a) Baseline: each dataset is treated as a trial. (b) Our system: decoupled scheduling.}
    \label{figure_system_eval}
\end{figure}


\subsection{Decoupled Scheduling for Evaluation}
\label{sub_sec_system_eval}
\noindent\textbf{Motivation}.
Evaluating the quality of LLMs based solely on a single metric, such as training loss, may not provide an accurate assessment \cite{HELM}. Therefore, it is vital to incorporate a variety of criteria and evaluate performance across an array of tasks \cite{LLMEvalSurvey}. Our LLM framework conducts regular evaluations for every checkpoint during the pretraining phase in our datacenter. This allows developers to track the progress of model training and identify the optimal model checkpoint. We aim for swift feedback to facilitate timely adjustment. However, as shown in Figure \ref{figure_cdf_job_types}, evaluation jobs experience the longest queuing delay due to limited resources and concurrent submission of numerous trials. Despite these challenges, we identify several opportunities to expedite the evaluation process.

\noindent\textbf{System Design}.
We develop a \emph{trial coordinator} to harmonize the operations of the cluster scheduler and LLM framework. This design incorporates the following three key techniques aimed at enhancing the efficiency of the evaluation process.

\noindent\textbf{\textsl{1. Decoupling Remote Model Loading}}.
Given the substantial size of LLMs, retrieving and loading them from remote storage can be a lengthy process. Furthermore, the concurrent execution of numerous evaluation tasks (around 60 datasets) can exacerbate this loading process due to increased contention. Figure \ref{figure_system_eval} (Left) shows the average model loading speed on a range of concurrent evaluation trials within \Scluster. It reveals a huge decline in loading speed when increasing the number of single-GPU trials from 1 to 8 on a single node, due to bandwidth limitation (25Gb/s) of our storage NIC. On the other hand, the loading speed stabilizes when the number of trials ranges from 8 to 256 GPUs.
This observation inspires us to take a strategic approach. Rather than submitting each evaluation dataset as a separate trial, we separated the model loading process from the evaluation process, as depicted in Figure \ref{figure_system_eval} (Right). Specifically, the trial coordinator initially retrieves the available node list from the cluster scheduler and then generates a series of precursor jobs for each node. These jobs load the model from remote storage to local shared memory. Then the coordinator submits the evaluation jobs to the scheduler, which loads the model via the high-bandwidth PCIe.
This method effectively utilizes spare host memory. After the evaluation finishes, the coordinator clears the files.


\noindent\textbf{\textsl{2. Decoupling Metric Computation}}.
As shown in Figure \ref{figure_profile_eval}, the evaluation process can often involve complex and time-consuming metric computation. For example, synthesized program correctness tests must be performed on coding datasets like HumanEval \cite{HumanEval} and MBPP \cite{MBPP}.
To address this issue, we decouple the metric computation process from the evaluation trial. After the model inference is performed on the GPU, its output is quickly saved into files, terminating the inference workload. Given that the outputs are typically text-based and thus small in size, this file-dumping process is swift. We then generate CPU jobs to carry out the metric computations. This approach effectively minimizes GPU idle time and accelerates the evaluation.

\noindent\textbf{\textsl{3. Prior-based Elastic Scheduling}}.
In addition to the decoupling approach, we notice that our prior knowledge regarding the approximate trial runtime for each evaluation dataset is quite robust. Furthermore, these datasets are flexible, allowing us to batch multiple sets into one trial to circumvent model loading. We can also break down large datasets and decouple metric computation. As a result, the trial coordinator can maximize GPU occupancy through decomposition, balance each GPU's workload using prior information, and employ a round-robin allocation strategy on sorted job queues. Moreover, we prioritize evaluation trials with lengthy CPU metric computations in the job queue to better overlap its computation. This approach not only enhances workload balance but also minimizes trial switch overhead.

\noindent\textbf{System Performance}.
We conducted a representative test of the trial coordinator using a typical evaluation job on a 7B size LLM, which involved evaluating the workload across 63 datasets. We measured the makespan necessary to complete all evaluation trials under two different conditions: a single node (representing limited resources) and four nodes (representing relatively ample resources). The trial coordinator can reduce the makespan by 1.3$\times$ and 1.8$\times$ respectively.

%% file: 7_Related.tex
\section{Discussion}
\label{sec_related_work}




\noindent\textbf{Related Work}. Due to the page limit, we provide a detailed discussion of our related work in Appendix \ref{appendix_related_work}.

\noindent\textbf{Scope Limitations}.
Despite our best efforts to analyze the workloads in \DC, it is an inescapable reality that we cannot cover all types of workloads. Limitations include: (1) Our analysis focuses on the developmental process preceding model serving and \DC does not encompass any serving jobs (i.e., workload in the deployment stage). (2) We concentrate our analysis predominantly on GPU jobs, providing limited room for CPU jobs. (3) We mainly characterize transformer-based, decoder-only architecture models (GPT-3 \cite{GPT-3} and LLaMA 2 \cite{LLaMA2}). For newer model architecture, we provide a simple characterization of the Mixture of Experts (MoE) model \cite{MoE} in Appendix \ref{appendix_subsec_moe}. Other model architectures like the Multimodal LLM \cite{CLIP} fall outside our scope of analysis.

\noindent\textbf{Continuous System Enhancement}.
With the rapid advancement of large models, the systems described in this work may not suffice for the demands of future workloads. In response, we are actively refining our system to accommodate advanced training workloads, including long sequence pretraining, MoE pretraining, and efficient RLHF.
Additionally, we are upgrading our infrastructure, with a particular focus on NIC, and expanding our computing cluster to facilitate larger-scale pretraining. Furthermore, we are exploring promising directions, such as improving the quality of LLMs through hyperparameter optimization using Hydro \cite{Hydro}, and providing efficient system support for emerging model architectures like Diffusion \cite{Diffusion} and Mamba \cite{Mamba}.

%% file: 8_Discussion.tex






%% file: 9_Conclusion.tex
\section{Conclusion}
\label{sec_conclusion}

In summary, we analyze LLM workloads and resource utilization in our datacenter \DC, revealing unique features and challenges of LLM development, such as resource inefficiencies and failure impacts. We also uncover potential opportunities to optimize systems tailored for LLMs and introduce our efforts for pretraining and evaluation workloads. We believe that our lessons and insights have broad applicability and can well benefit subsequent research.

%% file: 10_Appendix.tex
\newpage

\appendix

\begin{figure}[t]
    \centering
    \includegraphics[width=\linewidth]{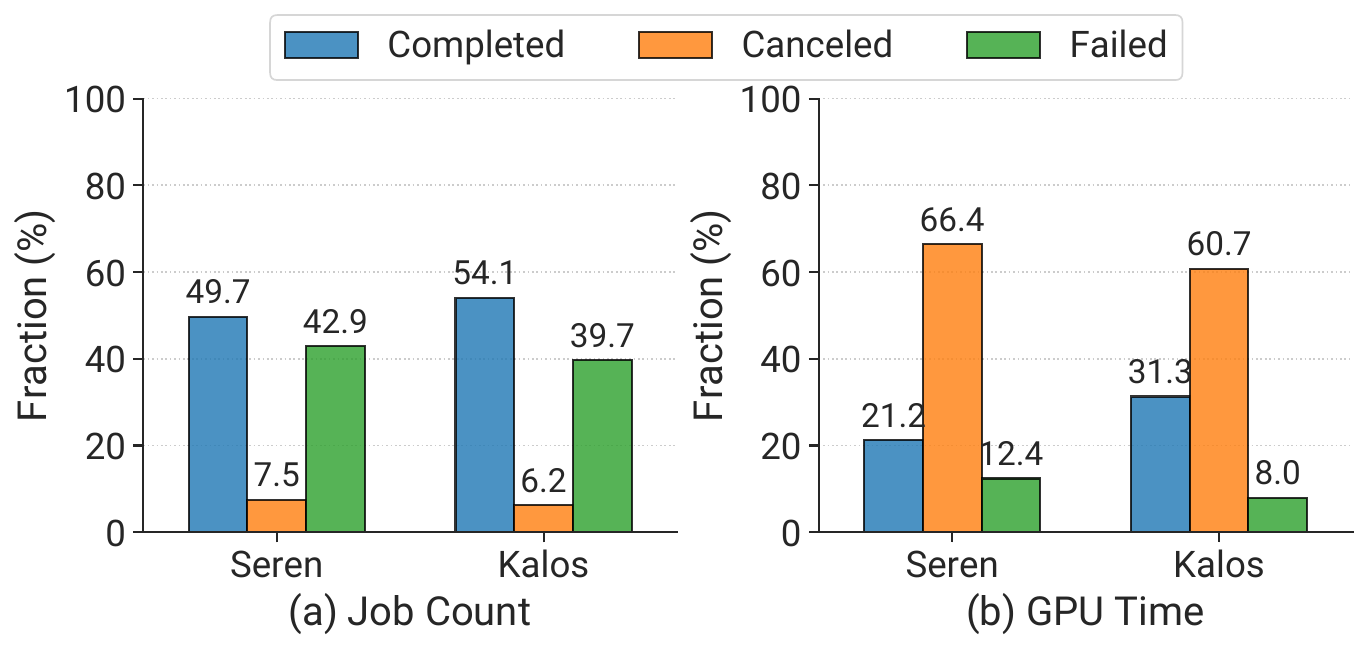}
    \caption{Final statuses of jobs in terms of (a) quantity and (b) utilized GPU resources.}
    \label{figure_job_state}
\end{figure}

\section{Supplementary Characterization}
\label{appendix_characterization}
In this section, we provide additional analysis to further characterize the workload features during our LLM development.

\subsection{Job Final Statuses}
\label{appendix_subsec_final_statuses}
\textbf{High Incompletion Rate}.
Figure \ref{figure_job_state} summarizes the distribution of three key final statuses across our two clusters, revealing a similar pattern. It is obvious that only approximately 20$\sim$30\% of resources are consumed by jobs that finally complete. Besides, about 40\% of jobs fail, utilizing 10\% of GPU resources. This suggests that failures predominantly occur in the early stages of execution, aligning with the statistics presented in Table \ref{table_failure_summary}. Canceled jobs, while constituting only around 7\% of the total job count, command over 60\% of GPU resources. This pattern suggests a prevalence of large-scale pretraining jobs being canceled by users. Beyond the common cancellation motives cited in prior studies (e.g., achieving desired model performance sooner than expected, early recognition of poor hyperparameter configuration) \cite{Helios,Gandiva,Optimus}, our experience with LLM pretraining has identified two additional frequent causes: (1) Users pausing jobs to adjust parameters in response to performance anomalies, such as loss spikes. (2) Jobs stalling due to infrastructure issues without throwing error messages, only to be addressed upon manual inspection by users, leading to significant resource wastage. These observations underscore the necessity for a failure-handling system that can autonomously detect and rectify faulty jobs, which is further elaborated in \S \ref{failure_recovery} and \S \ref{sub_sec_system_pretrain}.

\begin{figure}[t]
    \centering
    \includegraphics[width=\linewidth]{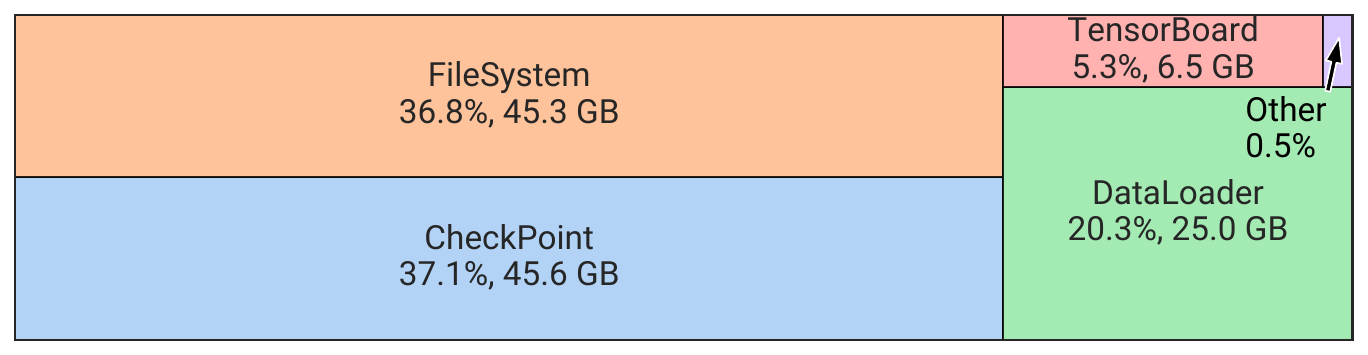}
    \caption{Distribution of host memory on a server in \Scluster during executing a pretraining job. Idle memory is not shown.}
    \label{figure_treemap_cpu_mem}
\end{figure}

\subsection{Host Memory}
\label{appendix_subsec_cpu_mem}
\textbf{Memory Footprint Breakdown}.
As depicted in Figure \ref{figure_treemap_cpu_mem}, we illustrate the distribution of active physical host memory within a compute node, which utilizes 123GB of the total 1TB available. This showcases a typical pattern of CPU memory consumption for pretraining jobs. However, it is important to note that memory usage can significantly vary across different tasks. Specifically, the memory footprint of the dataloader can be considerably larger when employing Megatron-LM \cite{MegatronLM}, which requires loading the metadata of the entire dataset. In contrast, our approach of loading data on-the-fly proves to be more memory-efficient without obviously impacting throughput. Furthermore, the memory requirements for asynchronous checkpointing (\S \ref{sub_sec_system_pretrain}) largely depend on the model size and training configurations. The memory footprint depicted in this figure corresponds to the configuration outlined in Figure \ref{figure_profile_pretrain}(a). In addition to the training processes, memory-intensive operations include TensorBoard \cite{TensorBoard} (6.5GB), the client daemon along the critical components (e.g., data and metadata caching) of the distributed file system (45.3GB). The remainder (0.6GB) encompasses Prometheus monitoring components, NVIDIA drivers, the Slurm scheduler daemon, and other system processes primarily related to sensor monitoring and system management. In general, there is a substantial amount of memory available, which can be leveraged for various purposes. Previous efforts \cite{tf.data,tf.data_service} have shown the potential for disaggregating CPU and memory usage from GPU allocations, suggesting that there may be additional optimization opportunities for LLMs, such as enhancing fault tolerance \cite{GEMINI}.

\begin{figure}[t]
    \centering
    \includegraphics[width=\linewidth]{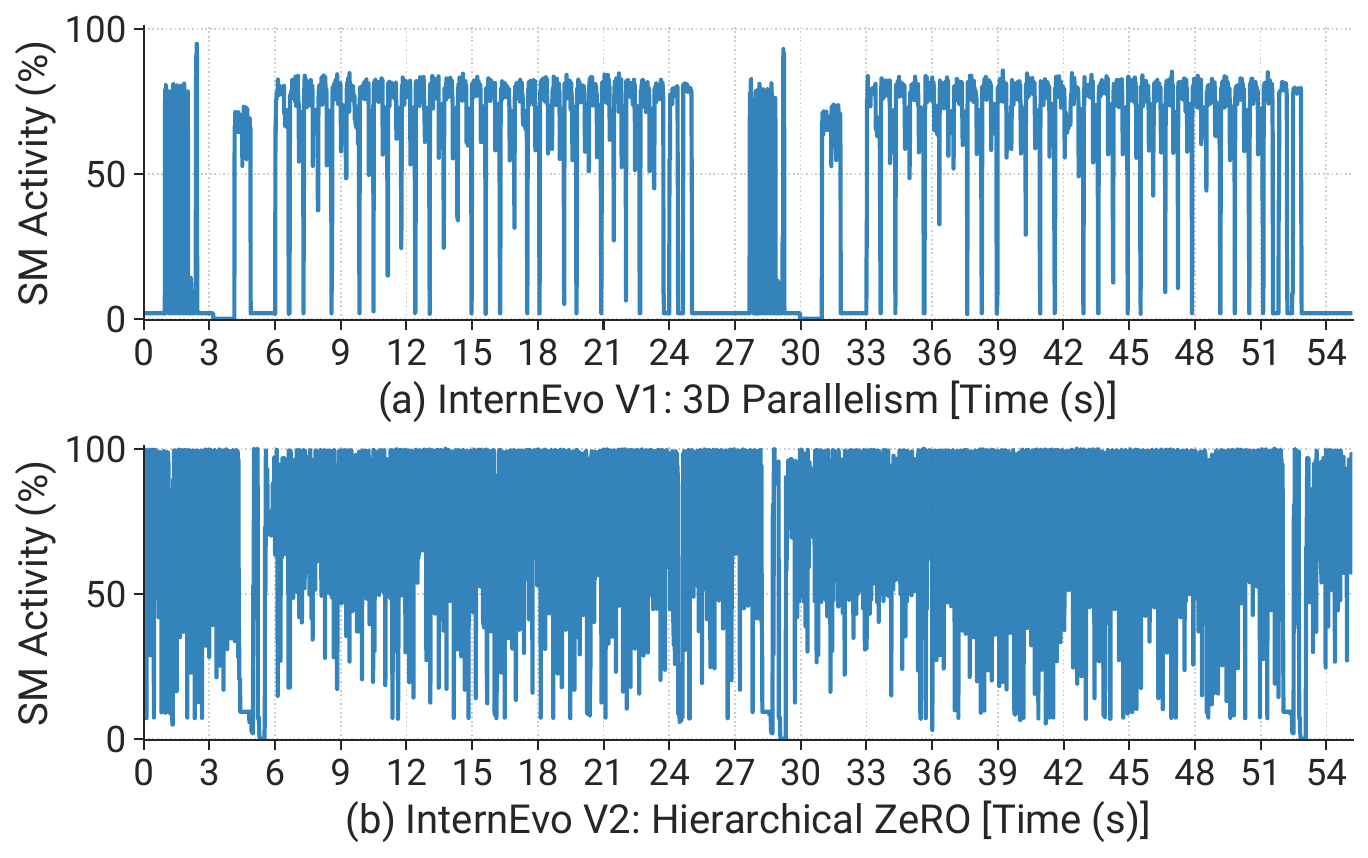}
    \caption{GPU SM utilization of pretraining a 123B LLM using different strategies of InternEvo \cite{InternEvo} over 1024 GPUs.}
    \label{figure_profile_pretrain_1024}
\end{figure}

\begin{figure}[t]
    \centering
    \includegraphics[width=\linewidth]{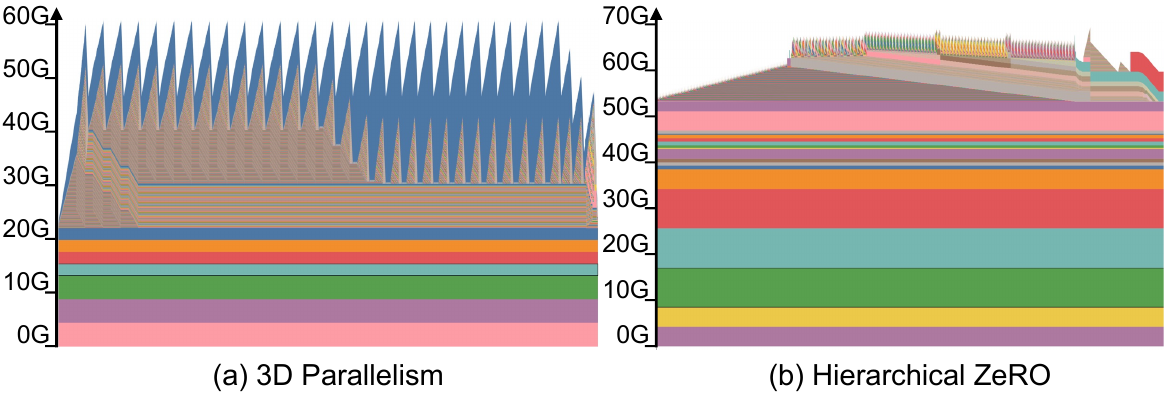}
    \caption{Memory snapshot under different pretraining strategies. Note that the extensive blue segment at the top of (a) is simplified and can be further broken down into massive fragments (memory allocations), similar to the lower part.}
    \label{figure_mem_timeline_1024}
\end{figure}

\subsection{Carbon Emission}
\label{appendix_subsec_carbon}
Our datacenter \DC has a PUE (Power Usage Effectiveness) of 1.25. Moreover, it operates on approximately 30.61\% of carbon-free energy (statistics for the year 2022), which includes renewable sources like solar and wind power and achieves a carbon emissions footprint (CO$_2$e) rate of 0.478 tCO$_2$e/MWh. Based on our node-level power consumption data, we calculate that \Scluster consumed approximately 673 MWh electricity in May 2023, which leads to total effective emissions of 321.7 tCO$_2$e. We believe that implementing advanced approaches like \cite{EnvPipe, CarbonExplorer, Zeus} can effectively reduce carbon emissions.

\subsection{Pretraining under Different Scale}
\label{appendix_subsec_pertraining}

Figures \ref{figure_profile_pretrain_1024} and \ref{figure_mem_timeline_1024} provide supplementary pretraining profiling results for a 123B LLM across 1024 GPUs. These figures complement the profiling observations depicted in Figures \ref{figure_profile_pretrain} and \ref{figure_mem_timeline}. It is evident that they present very similar patterns to the 2048 GPUs, demonstrating the generalizability of our characterization.

\begin{figure}[t]
    \centering
    \includegraphics[width=0.58\linewidth]{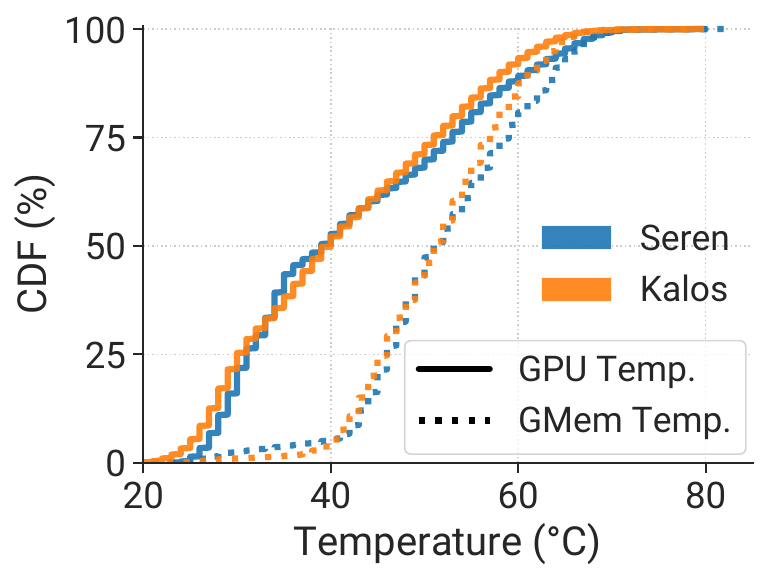}
    \caption{CDF of GPU core and GPU memory temperature.}
    \label{figure_temperature}
\end{figure}

\subsection{GPU Temperature}
\label{appendix_subsec_temperature}

Figure \ref{figure_temperature} depicts the temperature distributions of GPU core and GPU memory. The GPU memory temperature is generally higher than the GPU core temperature. As corresponding to power consumption distribution shown in Figure \ref{figure_power_cdf}, temperature presents a similar pattern in that some GPUs are under heavy load and have higher temperature (over 65$^\circ$C). These suggest a need for enhancements in our cluster's cooling system to address the issue of elevated temperatures.

\subsection{MoE Model}
\label{appendix_subsec_moe}

\begin{figure}[t]
    \centering
    \includegraphics[width=\linewidth]{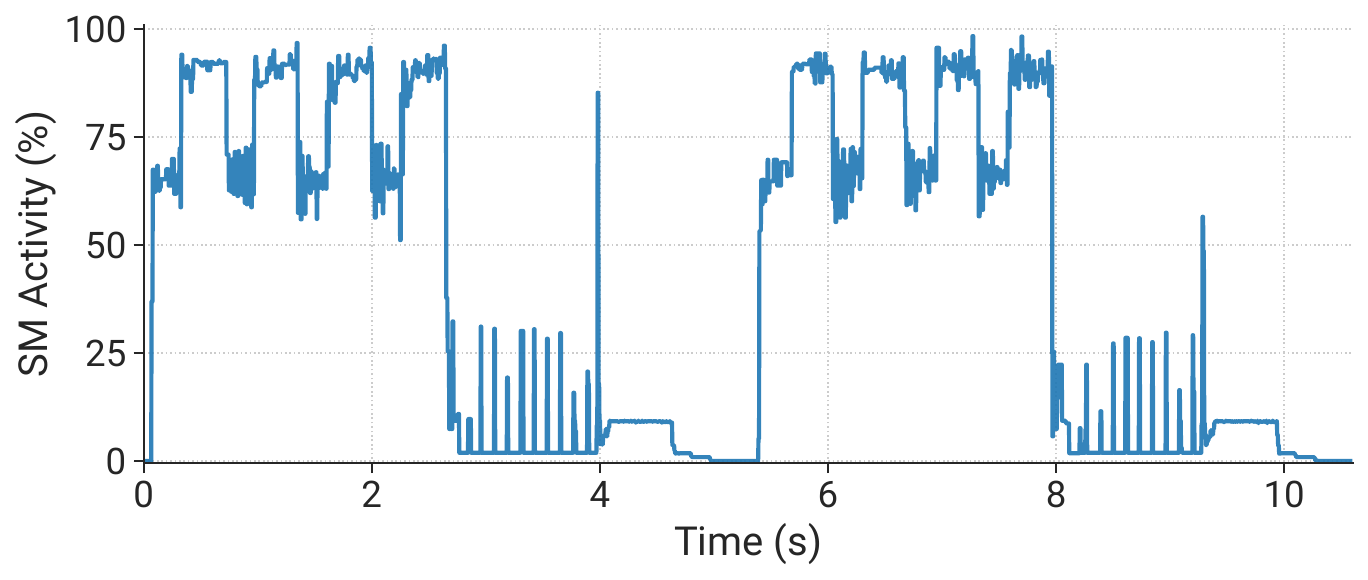}
    \caption{GPU SM utilization of pretraining a MoE model Mistral 7B \cite{Mistral7B} with 1024 GPUs in \Scluster.}
    \label{figure_MoE}
\end{figure}

\noindent As shown in Figure \ref{figure_MoE}, the MoE model presents much lower GPU utilization compared with the dense model shown in Figure \ref{figure_profile_pretrain}. This is mainly due to the fact that the MoE model requires frequent all-to-all communication and necessitates high-quality internode communication, however, our single IB NIC server cannot efficiently handle such job. Here we directly use the official training configuration released by Mistral. On the other hand, InternEvo is still under development and we are working on performing tailored optimizations for MoE models.

\section{Lessons of Troubleshooting}
\label{appendix_lessons}

\noindent\textbf{Slowdown Caused by Garbage Collection}. In LLM tasks programmed with Python as the interface language, irregular performance declines are sometimes observed, with severe instances causing the average throughput per step to decrease by 2-3 times. Through performance analysis using the Linux perf tool, we found that the critical function \texttt{list\_traverse} in garbage collection consumes 30\% of the execution time per step. This abnormal time consumption during garbage collection is typically caused by code defects. Apart from fixing these defects, we manually control the timing of garbage collection using Python's garbage collector interface in the code of InternEvo V2. We fix the garbage collection intervals for each rank, thereby avoiding the randomness timing that leads to serious performance issues.

\noindent\textbf{Memory Leakage Caused by Dataloader}. When a pretraining task runs for a considerably long duration, an error of \textsl{Dataloader Worker Killed} may emerge. This issue stems from a gradual memory leak caused by PyTorch's implementation of the dataloader when operated with \texttt{num\_worker>0}. This error occurs on average 27 hours after the start of a task. The root cause for this memory leak is the copy-on-write mechanism in the multiprocessing fork start method, coupled with a suboptimal design of the Python list. We avoid this issue by setting \texttt{num\_worker=0} and enabling garbage collection of the dataloder.

\section{Related Work}
\label{appendix_related_work}

\noindent\textbf{DL Workload Characterization}.
Prior works conduct DL workloads analysis from different companies. Philly \cite{Philly} provides insights on the impact of job locality on queuing delay and resource utilization from Microsoft, in addition to identifying different failure reasons. Helios \cite{Helios} from SenseTime illustrates the nature of cluster resource utilization and user disparity, evaluated from the perspectives of the cluster, job, and user. Alibaba's PAI \cite{MLaaS} contributes to this discourse by analyzing the challenges encountered within their clusters from both temporal and spatial viewpoints. Different from them, we focus on the characteristics of LLM workloads. Concurrently, MegaScale \cite{MegaScale} presents ByteDance's experience in training LLMs using a formidable array of over 10,000 GPUs, complementing our focus with their practical insights.

\noindent\textbf{Fault Tolerance Systems}.
Fault tolerance is a crucial consideration across various disciplines. In the context of LLM systems, Varuna \cite{Varuna}, Bamboo \cite{Bamboo}, and Oobleck \cite{Oobleck} focus on the fast recovery from failures in cloud spot instance scenarios. Gemini \cite{GEMINI} facilitates swift recovery through CPU memory checkpointing. In addition, a body of research works dedicated to GPU-related failure analysis \cite{GPULifetimeSC20, GPUerrorHPCA15, GPUReliabilitySC15, GPUerrorDSN21, NetworkFailures, Philly, DLFailures, CPR}, while several deep learning schedulers \cite{AFS, Lyra, SHEPHERD} consider fault tolerance. Furthermore, a series of studies have profiled \cite{Collie, PCIeCongestion, Justitia} or diagnosed \cite{Hostping, Pingmesh, Deepview, NetBouncer, SNAP, MegaScale} potential performance bottlenecks within RDMA or intra-host network communication.
We provide a thorough analysis of the failure events in LLM workloads and propose a LLM-involved diagnosis system.

\section{Resource Links}
\label{appendix_artifact}
InternLM is a project focusing on LLM research in Shanghai AI Laboratory. InternLM team keeps open-sourcing high-quality LLMs as well as a full-stack toolchain for LLM development. More resources can be accessed via the following links:

\begin{tcolorbox}[boxrule=0pt, left=1mm, right=1mm, top=0.5mm, bottom=0.5mm, arc=1mm, title=\textbf{InternLM Links}]
    \textbf{Project}: \url{https://github.com/InternLM}

    \textbf{Trace}: \url{https://github.com/InternLM/AcmeTrace}

    \textbf{System}: \url{https://github.com/InternLM/InternEvo}

    \textbf{Model}: \url{https://huggingface.co/internlm}
\end{tcolorbox}

%% file: nsdi24.bbl
\begin{thebibliography}{100}

\bibitem{BLOOM}
Bloom.
\newblock \url{https://bigscience.huggingface.co/blog/bloom}, 2024.

\bibitem{ChatGPT}
Chatgpt.
\newblock \url{https://openai.com/blog/chatgpt}, 2024.

\bibitem{DLRover}
Dlrover: An automatic distributed deep learning system.
\newblock \url{https://github.com/intelligent-machine-learning/dlrover}, 2024.

\bibitem{GitHubCopilot}
Github copilot.
\newblock \url{https://github.com/features/copilot/}, 2024.

\bibitem{InfiniBand}
Infiniband networking.
\newblock \url{https://www.nvidia.com/en-us/networking/products/infiniband/}, 2024.

\bibitem{A100}
Nvidia a100 tensor core gpu.
\newblock \url{https://www.nvidia.com/en-us/data-center/a100/}, 2024.

\bibitem{nvdcgm}
Nvidia data center gpu manager.
\newblock \url{https://developer.nvidia.com/dcgm}, 2024.

\bibitem{nvsmi}
Nvidia-smi.
\newblock \url{https://developer.nvidia.com/nvidia-system-management-interface}, 2024.

\bibitem{nvlink}
Nvlink and nvswitch.
\newblock \url{https://www.nvidia.com/en-us/data-center/nvlink/}, 2024.

\bibitem{AMP}
Pytorch automatic mixed precision training.
\newblock \url{https://pytorch.org/docs/stable/amp}, 2024.

\bibitem{PytorchMemShot}
Pytorch memory snapshottool.
\newblock \url{https://pytorch.org/blog/understanding-gpu-memory-1}, 2024.

\bibitem{ipmi}
Supermicro ipmi.
\newblock \url{https://www.supermicro.com/en/solutions/management-software/ipmi-utilities}, 2024.

\bibitem{TensorBoard}
Tensorboard.
\newblock \url{https://www.tensorflow.org/tensorboard}, 2024.

\bibitem{CarbonExplorer}
Bilge Acun, Benjamin Lee, Fiodar Kazhamiaka, Kiwan Maeng, Udit Gupta, Manoj Chakkaravarthy, David Brooks, and Carole-Jean Wu.
\newblock Carbon explorer: A holistic framework for designing carbon aware datacenters.
\newblock In {\em Proceedings of the 28th ACM International Conference on Architectural Support for Programming Languages and Operating Systems}, ASPLOS '23, 2023.

\bibitem{Varuna}
Sanjith Athlur, Nitika Saran, Muthian Sivathanu, Ramachandran Ramjee, and Nipun Kwatra.
\newblock Varuna: scalable, low-cost training of massive deep learning models.
\newblock In {\em Proceedings of the Seventeenth European Conference on Computer Systems}, EuroSys '22, 2022.

\bibitem{tf.data_service}
Andrew Audibert, Yang Chen, Dan Graur, Ana Klimovic, Ji\v{r}\'{\i} \v{S}im\v{s}a, and Chandramohan~A. Thekkath.
\newblock tf.data service: A case for disaggregating ml input data processing.
\newblock In {\em Proceedings of the ACM Symposium on Cloud Computing}, SoCC '23, 2023.

\bibitem{MBPP}
Jacob Austin, Augustus Odena, Maxwell Nye, Maarten Bosma, Henryk Michalewski, David Dohan, Ellen Jiang, Carrie Cai, Michael Terry, Quoc Le, and Charles Sutton.
\newblock Program synthesis with large language models.
\newblock {\em CoRR}, 2021.

\bibitem{LSTM}
Dzmitry Bahdanau, Kyunghyun Cho, and Yoshua Bengio.
\newblock Neural machine translation by jointly learning to align and translate.
\newblock In {\em 3rd International Conference on Learning Representations}, ICLR '15, 2015.

\bibitem{FoundationModel}
Rishi Bommasani, Drew~A. Hudson, Ehsan Adeli, Russ Altman, Simran Arora, Sydney von Arx, Michael~S. Bernstein, Jeannette Bohg, Antoine Bosselut, Emma Brunskill, Erik Brynjolfsson, Shyamal Buch, Dallas Card, Rodrigo Castellon, Niladri Chatterji, Annie Chen, Kathleen Creel, Jared~Quincy Davis, Dora Demszky, Chris Donahue, Moussa Doumbouya, Esin Durmus, Stefano Ermon, John Etchemendy, Kawin Ethayarajh, Li~Fei-Fei, Chelsea Finn, Trevor Gale, Lauren Gillespie, Karan Goel, Noah Goodman, Shelby Grossman, Neel Guha, Tatsunori Hashimoto, Peter Henderson, John Hewitt, Daniel~E. Ho, Jenny Hong, Kyle Hsu, Jing Huang, Thomas Icard, Saahil Jain, Dan Jurafsky, Pratyusha Kalluri, Siddharth Karamcheti, Geoff Keeling, Fereshte Khani, Omar Khattab, Pang~Wei Koh, Mark Krass, Ranjay Krishna, Rohith Kuditipudi, Ananya Kumar, Faisal Ladhak, Mina Lee, Tony Lee, Jure Leskovec, Isabelle Levent, Xiang~Lisa Li, Xuechen Li, Tengyu Ma, Ali Malik, Christopher~D. Manning, Suvir Mirchandani, Eric Mitchell, Zanele Munyikwa, Suraj Nair, Avanika Narayan, Deepak Narayanan, Ben Newman, Allen Nie, Juan~Carlos Niebles, Hamed Nilforoshan, Julian Nyarko, Giray Ogut, Laurel Orr, Isabel Papadimitriou, Joon~Sung Park, Chris Piech, Eva Portelance, Christopher Potts, Aditi Raghunathan, Rob Reich, Hongyu Ren, Frieda Rong, Yusuf Roohani, Camilo Ruiz, Jack Ryan, Christopher Ré, Dorsa Sadigh, Shiori Sagawa, Keshav Santhanam, Andy Shih, Krishnan Srinivasan, Alex Tamkin, Rohan Taori, Armin~W. Thomas, Florian Tramèr, Rose~E. Wang, William Wang, Bohan Wu, Jiajun Wu, Yuhuai Wu, Sang~Michael Xie, Michihiro Yasunaga, Jiaxuan You, Matei Zaharia, Michael Zhang, Tianyi Zhang, Xikun Zhang, Yuhui Zhang, Lucia Zheng, Kaitlyn Zhou, and Percy Liang.
\newblock On the opportunities and risks of foundation models.
\newblock {\em CoRR}, 2021.

\bibitem{GPT-3}
Tom Brown, Benjamin Mann, Nick Ryder, Melanie Subbiah, Jared~D Kaplan, Prafulla Dhariwal, Arvind Neelakantan, Pranav Shyam, Girish Sastry, Amanda Askell, Sandhini Agarwal, Ariel Herbert-Voss, Gretchen Krueger, Tom Henighan, Rewon Child, Aditya Ramesh, Daniel Ziegler, Jeffrey Wu, Clemens Winter, Chris Hesse, Mark Chen, Eric Sigler, Mateusz Litwin, Scott Gray, Benjamin Chess, Jack Clark, Christopher Berner, Sam McCandlish, Alec Radford, Ilya Sutskever, and Dario Amodei.
\newblock Language models are few-shot learners.
\newblock In {\em Advances in Neural Information Processing Systems}, NeurIPS '20, 2020.

\bibitem{BorgK8S}
Brendan Burns, Brian Grant, David Oppenheimer, Eric Brewer, and John Wilkes.
\newblock Borg, omega, and kubernetes: Lessons learned from three container-management systems over a decade.
\newblock {\em Queue}, 2016.

\bibitem{LLMEvalSurvey}
Yupeng Chang, Xu~Wang, Jindong Wang, Yuan Wu, Linyi Yang, Kaijie Zhu, Hao Chen, Xiaoyuan Yi, Cunxiang Wang, Yidong Wang, Wei Ye, Yue Zhang, Yi~Chang, Philip~S. Yu, Qiang Yang, and Xing Xie.
\newblock A survey on evaluation of large language models.
\newblock {\em CoRR}, 2023.

\bibitem{LogFailures}
An~Ran Chen.
\newblock An empirical study on leveraging logs for debugging production failures.
\newblock In {\em Proceedings of the 41st International Conference on Software Engineering: Companion Proceedings}, ICSE '19, 2019.

\bibitem{HumanEval}
Mark Chen, Jerry Tworek, Heewoo Jun, Qiming Yuan, Henrique~Ponde de~Oliveira~Pinto, Jared Kaplan, Harri Edwards, Yuri Burda, Nicholas Joseph, Greg Brockman, Alex Ray, Raul Puri, Gretchen Krueger, Michael Petrov, Heidy Khlaaf, Girish Sastry, Pamela Mishkin, Brooke Chan, Scott Gray, Nick Ryder, Mikhail Pavlov, Alethea Power, Lukasz Kaiser, Mohammad Bavarian, Clemens Winter, Philippe Tillet, Felipe~Petroski Such, Dave Cummings, Matthias Plappert, Fotios Chantzis, Elizabeth Barnes, Ariel Herbert-Voss, William~Hebgen Guss, Alex Nichol, Alex Paino, Nikolas Tezak, Jie Tang, Igor Babuschkin, Suchir Balaji, Shantanu Jain, William Saunders, Christopher Hesse, Andrew~N. Carr, Jan Leike, Josh Achiam, Vedant Misra, Evan Morikawa, Alec Radford, Matthew Knight, Miles Brundage, Mira Murati, Katie Mayer, Peter Welinder, Bob McGrew, Dario Amodei, Sam McCandlish, Ilya Sutskever, and Wojciech Zaremba.
\newblock Evaluating large language models trained on code.
\newblock {\em CoRR}, 2021.

\bibitem{InternEvo}
Qiaoling Chen, Diandian Gu, Guoteng Wang, Xun Chen, YingTong Xiong, Ting Huang, Qinghao Hu, Xin Jin, Yonggang Wen, Tianwei Zhang, and Peng Sun.
\newblock Internevo: Efficient long-sequence large language model training via hybrid parallelism and redundant sharding.
\newblock {\em CoRR}, abs/2401.09149, 2024.

\bibitem{EnvPipe}
Sangjin Choi, Inhoe Koo, Jeongseob Ahn, Myeongjae Jeon, and Youngjin Kwon.
\newblock Envpipe: Performance-preserving dnn training framework for saving energy.
\newblock In {\em 2023 {USENIX} Annual Technical Conference}, {USENIX} {ATC} '23, 2023.

\bibitem{PaLM}
Aakanksha Chowdhery, Sharan Narang, Jacob Devlin, Maarten Bosma, Gaurav Mishra, Adam Roberts, Paul Barham, Hyung~Won Chung, Charles Sutton, Sebastian Gehrmann, Parker Schuh, Kensen Shi, Sasha Tsvyashchenko, Joshua Maynez, Abhishek Rao, Parker Barnes, Yi~Tay, Noam Shazeer, Vinodkumar Prabhakaran, Emily Reif, Nan Du, Ben Hutchinson, Reiner Pope, James Bradbury, Jacob Austin, Michael Isard, Guy Gur-Ari, Pengcheng Yin, Toju Duke, Anselm Levskaya, Sanjay Ghemawat, Sunipa Dev, Henryk Michalewski, Xavier Garcia, Vedant Misra, Kevin Robinson, Liam Fedus, Denny Zhou, Daphne Ippolito, David Luan, Hyeontaek Lim, Barret Zoph, Alexander Spiridonov, Ryan Sepassi, David Dohan, Shivani Agrawal, Mark Omernick, Andrew~M. Dai, Thanumalayan~Sankaranarayana Pillai, Marie Pellat, Aitor Lewkowycz, Erica Moreira, Rewon Child, Oleksandr Polozov, Katherine Lee, Zongwei Zhou, Xuezhi Wang, Brennan Saeta, Mark Diaz, Orhan Firat, Michele Catasta, Jason Wei, Kathy Meier-Hellstern, Douglas Eck, Jeff Dean, Slav Petrov, and Noah Fiedel.
\newblock Palm: Scaling language modeling with pathways.
\newblock {\em CoRR}, 2022.

\bibitem{FlashAttention2}
Tri Dao.
\newblock Flashattention-2: Faster attention with better parallelism and work partitioning.
\newblock {\em CoRR}, 2023.

\bibitem{FlashAttention}
Tri Dao, Daniel~Y Fu, Stefano Ermon, Atri Rudra, and Christopher Re.
\newblock Flashattention: Fast and memory-efficient exact attention with io-awareness.
\newblock In {\em Advances in Neural Information Processing Systems}, NeurIPS '22, 2022.

\bibitem{LLMint8}
Tim Dettmers, Mike Lewis, Younes Belkada, and Luke Zettlemoyer.
\newblock Llm.int8(): 8-bit matrix multiplication for transformers at scale.
\newblock In {\em Advances in Neural Information Processing Systems}, NeurIPS '22, 2022.

\bibitem{BERT}
Jacob Devlin, Ming-Wei Chang, Kenton Lee, and Kristina Toutanova.
\newblock {BERT}: Pre-training of deep bidirectional transformers for language understanding.
\newblock In {\em Proceedings of the 2019 Conference of the North American Chapter of the Association for Computational Linguistics}, NAACL '19, 2019.

\bibitem{Check-N-Run}
Assaf Eisenman, Kiran~Kumar Matam, Steven Ingram, Dheevatsa Mudigere, Raghuraman Krishnamoorthi, Krishnakumar Nair, Misha Smelyanskiy, and Murali Annavaram.
\newblock {Check-N-Run}: a checkpointing system for training deep learning recommendation models.
\newblock In {\em 19th USENIX Symposium on Networked Systems Design and Implementation}, NSDI '22, 2022.

\bibitem{NetworkFailures}
Phillipa Gill, Navendu Jain, and Nachiappan Nagappan.
\newblock Understanding network failures in data centers: measurement, analysis, and implications.
\newblock In {\em Proceedings of the Annual Conference of the ACM Special Interest Group on Data Communication}, SIGCOMM '11, 2011.

\bibitem{Mamba}
Albert Gu and Tri Dao.
\newblock Mamba: Linear-time sequence modeling with selective state spaces.
\newblock {\em CoRR}, abs/2312.00752, 2023.

\bibitem{Tiresias}
Juncheng Gu, Mosharaf Chowdhury, Kang~G. Shin, Yibo Zhu, Myeongjae Jeon, Junjie Qian, Hongqiang Liu, and Chuanxiong Guo.
\newblock Tiresias: A {GPU} cluster manager for distributed deep learning.
\newblock In {\em 16th {USENIX} Symposium on Networked Systems Design and Implementation}, NSDI '19, 2019.

\bibitem{Pingmesh}
Chuanxiong Guo, Lihua Yuan, Dong Xiang, Yingnong Dang, Ray Huang, Dave Maltz, Zhaoyi Liu, Vin Wang, Bin Pang, Hua Chen, Zhi-Wei Lin, and Varugis Kurien.
\newblock Pingmesh: A large-scale system for data center network latency measurement and analysis.
\newblock In {\em Proceedings of the Annual Conference of the ACM Special Interest Group on Data Communication}, SIGCOMM '15, 2015.

\bibitem{LoRA}
Edward~J Hu, yelong shen, Phillip Wallis, Zeyuan Allen-Zhu, Yuanzhi Li, Shean Wang, Lu~Wang, and Weizhu Chen.
\newblock Lora: Low-rank adaptation of large language models.
\newblock In {\em International Conference on Learning Representations}, ICLR '22, 2022.

\bibitem{Helios}
Qinghao Hu, Peng Sun, Shengen Yan, Yonggang Wen, and Tianwei Zhang.
\newblock Characterization and prediction of deep learning workloads in large-scale gpu datacenters.
\newblock In {\em Proceedings of the International Conference for High Performance Computing, Networking, Storage and Analysis}, SC '21, 2021.

\bibitem{Hydro}
Qinghao Hu, Zhisheng Ye, Meng Zhang, Qiaoling Chen, Peng Sun, Yonggang Wen, and Tianwei Zhang.
\newblock Hydro: {Surrogate-Based} hyperparameter tuning service in datacenters.
\newblock In {\em 17th USENIX Symposium on Operating Systems Design and Implementation}, OSDI '23, 2023.

\bibitem{Lucid}
Qinghao Hu, Meng Zhang, Peng Sun, Yonggang Wen, and Tianwei Zhang.
\newblock Lucid: A non-intrusive, scalable and interpretable scheduler for deep learning training jobs.
\newblock In {\em Proceedings of the 28th International Conference on Architectural Support for Programming Languages and Operating Systems}, ASPLOS '23, 2023.

\bibitem{Metastable}
Lexiang Huang, Matthew Magnusson, Abishek~Bangalore Muralikrishna, Salman Estyak, Rebecca Isaacs, Abutalib Aghayev, Timothy Zhu, and Aleksey Charapko.
\newblock Metastable failures in the wild.
\newblock In {\em 16th USENIX Symposium on Operating Systems Design and Implementation}, OSDI '22, 2022.

\bibitem{AFS}
Changho Hwang, Taehyun Kim, Sunghyun Kim, Jinwoo Shin, and KyoungSoo Park.
\newblock Elastic resource sharing for distributed deep learning.
\newblock In {\em 18th {USENIX} Symposium on Networked Systems Design and Implementation}, NSDI '21, 2021.

\bibitem{DataMovement}
Andrei Ivanov, Nikoli Dryden, Tal Ben-Nun, Shigang Li, and Torsten Hoefler.
\newblock Data movement is all you need: A case study on optimizing transformers.
\newblock In {\em Proceedings of Machine Learning and Systems}, MLSys '21, 2021.

\bibitem{Oobleck}
Insu Jang, Zhenning Yang, Zhen Zhang, Xin Jin, and Mosharaf Chowdhury.
\newblock Oobleck: Resilient distributed training of large models using pipeline templates.
\newblock In {\em Proceedings of the ACM SIGOPS 29th Symposium on Operating Systems Principles}, SOSP '23, 2023.

\bibitem{Philly}
Myeongjae Jeon, Shivaram Venkataraman, Amar Phanishayee, Junjie Qian, Wencong Xiao, and Fan Yang.
\newblock Analysis of large-scale multi-tenant {GPU} clusters for {DNN} training workloads.
\newblock In {\em 2019 {USENIX} Annual Technical Conference}, {USENIX} {ATC} '19, 2019.

\bibitem{Mistral7B}
Albert~Q. Jiang, Alexandre Sablayrolles, Arthur Mensch, Chris Bamford, Devendra~Singh Chaplot, Diego de~las Casas, Florian Bressand, Gianna Lengyel, Guillaume Lample, Lucile Saulnier, Lélio~Renard Lavaud, Marie-Anne Lachaux, Pierre Stock, Teven~Le Scao, Thibaut Lavril, Thomas Wang, Timothée Lacroix, and William~El Sayed.
\newblock Mistral 7b.
\newblock {\em CoRR}, abs/2310.06825, 2023.

\bibitem{MegaScale}
Ziheng Jiang, Haibin Lin, Yinmin Zhong, Qi~Huang, Yangrui Chen, Zhi Zhang, Yanghua Peng, Xiang Li, Cong Xie, Shibiao Nong, Yulu Jia, Sun He, Hongmin Chen, Zhihao Bai, Qi~Hou, Shipeng Yan, Ding Zhou, Yiyao Sheng, Zhuo Jiang, Haohan Xu, Haoran Wei, Zhang Zhang, Pengfei Nie, Leqi Zou, Sida Zhao, Liang Xiang, Zherui Liu, Zhe Li, Xiaoying Jia, Jianxi Ye, Xin Jin, and Xin Liu.
\newblock Megascale: Scaling large language model training to more than 10,000 gpus.
\newblock {\em CoRR}, abs/2402.15627, 2024.

\bibitem{Adam}
Diederik~P Kingma and Jimmy Ba.
\newblock Adam: A method for stochastic optimization.
\newblock In {\em International Conference on Learning Representations}, ICLR '15, 2015.

\bibitem{Collie}
Xinhao Kong, Yibo Zhu, Huaping Zhou, Zhuo Jiang, Jianxi Ye, Chuanxiong Guo, and Danyang Zhuo.
\newblock Collie: Finding performance anomalies in {RDMA} subsystems.
\newblock In {\em 19th USENIX Symposium on Networked Systems Design and Implementation}, NSDI '22, 2022.

\bibitem{Megatron-LMV3}
Vijay Korthikanti, Jared Casper, Sangkug Lym, Lawrence McAfee, Michael Andersch, Mohammad Shoeybi, and Bryan Catanzaro.
\newblock Reducing activation recomputation in large transformer models.
\newblock {\em CoRR}, 2022.

\bibitem{vLLM}
Woosuk Kwon, Zhuohan Li, Siyuan Zhuang, Ying Sheng, Lianmin Zheng, Cody~Hao Yu, Joseph Gonzalez, Hao Zhang, and Ion Stoica.
\newblock Efficient memory management for large language model serving with pagedattention.
\newblock In {\em Proceedings of the ACM SIGOPS 29th Symposium on Operating Systems Principles}, SOSP '23, 2023.

\bibitem{LogParsing}
Van-Hoang Le and Hongyu Zhang.
\newblock Log parsing: How far can chatgpt go?
\newblock In {\em Proceedings of IEEE/ACM International Conference on Automated Software Engineering}, ASE '23, 2023.

\bibitem{LogPPT}
Van-Hoang Le and Hongyu Zhang.
\newblock Log parsing with prompt-based few-shot learning.
\newblock In {\em Proceedings of the 45th International Conference on Software Engineering}, ICSE '23, 2023.

\bibitem{CNN}
Yann LeCun, L{\'e}on Bottou, Yoshua Bengio, and Patrick Haffner.
\newblock Gradient-based learning applied to document recognition.
\newblock {\em Proceedings of the IEEE}, 1998.

\bibitem{RAG}
Patrick Lewis, Ethan Perez, Aleksandra Piktus, Fabio Petroni, Vladimir Karpukhin, Naman Goyal, Heinrich Küttler, Mike Lewis, Wen-tau Yih, Tim Rocktäschel, Sebastian Riedel, and Douwe Kiela.
\newblock Retrieval-augmented generation for knowledge-intensive nlp tasks.
\newblock In {\em Advances in Neural Information Processing Systems}, NeurIPS '20, 2020.

\bibitem{Lyra}
Jiamin Li, Hong Xu, Yibo Zhu, Zherui Liu, Chuanxiong Guo, and Cong Wang.
\newblock Lyra: Elastic scheduling for deep learning clusters.
\newblock In {\em Proceedings of the Eighteenth European Conference on Computer Systems}, EuroSys '23, 2023.

\bibitem{AlpaServe}
Zhuohan Li, Lianmin Zheng, Yinmin Zhong, Vincent Liu, Ying Sheng, Xin Jin, Yanping Huang, Zhifeng Chen, Hao Zhang, Joseph~E. Gonzalez, and Ion Stoica.
\newblock {AlpaServe}: Statistical multiplexing with model parallelism for deep learning serving.
\newblock In {\em 17th USENIX Symposium on Operating Systems Design and Implementation}, OSDI '23, 2023.

\bibitem{HELM}
Percy Liang, Rishi Bommasani, Tony Lee, Dimitris Tsipras, Dilara Soylu, Michihiro Yasunaga, Yian Zhang, Deepak Narayanan, Yuhuai Wu, Ananya Kumar, Benjamin Newman, Binhang Yuan, Bobby Yan, Ce~Zhang, Christian Cosgrove, Christopher~D. Manning, Christopher Ré, Diana Acosta-Navas, Drew~A. Hudson, Eric Zelikman, Esin Durmus, Faisal Ladhak, Frieda Rong, Hongyu Ren, Huaxiu Yao, Jue Wang, Keshav Santhanam, Laurel Orr, Lucia Zheng, Mert Yuksekgonul, Mirac Suzgun, Nathan Kim, Neel Guha, Niladri Chatterji, Omar Khattab, Peter Henderson, Qian Huang, Ryan Chi, Sang~Michael Xie, Shibani Santurkar, Surya Ganguli, Tatsunori Hashimoto, Thomas Icard, Tianyi Zhang, Vishrav Chaudhary, William Wang, Xuechen Li, Yifan Mai, Yuhui Zhang, and Yuta Koreeda.
\newblock Holistic evaluation of language models.
\newblock {\em CoRR}, 2022.

\bibitem{Hostping}
Kefei Liu, Zhuo Jiang, Jiao Zhang, Haoran Wei, Xiaolong Zhong, Lizhuang Tan, Tian Pan, and Tao Huang.
\newblock Hostping: Diagnosing intra-host network bottlenecks in {RDMA} servers.
\newblock In {\em 20th USENIX Symposium on Networked Systems Design and Implementation}, NSDI '23, 2023.

\bibitem{CPR}
Kiwan Maeng, Shivam Bharuka, Isabel Gao, Mark Jeffrey, Vikram Saraph, Bor-Yiing Su, Caroline Trippel, Jiyan Yang, Mike Rabbat, Brandon Lucia, and Carole-Jean Wu.
\newblock Understanding and improving failure tolerant training for deep learning recommendation with partial recovery.
\newblock In {\em Proceedings of Machine Learning and Systems}, MLSys '21, 2021.

\bibitem{Themis}
Kshiteej Mahajan, Arjun Balasubramanian, Arjun Singhvi, Shivaram Venkataraman, Aditya Akella, Amar Phanishayee, and Shuchi Chawla.
\newblock Themis: Fair and efficient {GPU} cluster scheduling.
\newblock In {\em 17th {USENIX} Symposium on Networked Systems Design and Implementation}, NSDI '20, 2020.

\bibitem{PCIeCongestion}
Maxime Martinasso, Grzegorz Kwasniewski, Sadaf~R. Alam, Thomas~C. Schulthess, and Torsten Hoefler.
\newblock A pcie congestion-aware performance model for densely populated accelerator servers.
\newblock In {\em Proceedings of the International Conference for High Performance Computing, Networking, Storage and Analysis}, 2016.

\bibitem{HottestMonth}
Jackie McGuinness and Katherine Rohloff.
\newblock Nasa clocks july 2023 as hottest month on record ever since 1880.
\newblock https://www.nasa.gov/news-release/nasa-clocks-july-2023-as-hottest-month-on-record-ever-since-1880/, 2024.

\bibitem{CheckFreq}
Jayashree Mohan, Amar Phanishayee, and Vijay Chidambaram.
\newblock {CheckFreq}: Frequent, {Fine-Grained} {DNN} checkpointing.
\newblock In {\em 19th USENIX Conference on File and Storage Technologies}, FAST '21, 2021.

\bibitem{tf.data}
Derek~Gordon Murray, Jiri Simsa, Ana Klimovic, and Ihor Indyk.
\newblock tf.data: A machine learning data processing framework.
\newblock {\em Proceedings of the VLDB Endowment}, 2021.

\bibitem{DISTALYZER}
Karthik Nagaraj, Charles Killian, and Jennifer Neville.
\newblock Structured comparative analysis of systems logs to diagnose performance problems.
\newblock In {\em 9th USENIX Symposium on Networked Systems Design and Implementation}, NSDI '12, 2012.

\bibitem{PipeDream}
Deepak Narayanan, Aaron Harlap, Amar Phanishayee, Vivek Seshadri, Nikhil~R. Devanur, Gregory~R. Ganger, Phillip~B. Gibbons, and Matei Zaharia.
\newblock Pipedream: generalized pipeline parallelism for dnn training.
\newblock In {\em Proceedings of the 27th ACM Symposium on Operating Systems Principles}, SOSP '19, 2019.

\bibitem{MegatronLM}
Deepak Narayanan, Mohammad Shoeybi, Jared Casper, Patrick LeGresley, Mostofa Patwary, Vijay Korthikanti, Dmitri Vainbrand, Prethvi Kashinkunti, Julie Bernauer, Bryan Catanzaro, Amar Phanishayee, and Matei Zaharia.
\newblock Efficient large-scale language model training on gpu clusters using megatron-lm.
\newblock In {\em Proceedings of the International Conference for High Performance Computing, Networking, Storage and Analysis}, SC '21, 2021.

\bibitem{DeepFreeze}
Bogdan Nicolae, Jiali Li, Justin~M Wozniak, George Bosilca, Matthieu Dorier, and Franck Cappello.
\newblock Deepfreeze: Towards scalable asynchronous checkpointing of deep learning models.
\newblock In {\em 2020 20th IEEE/ACM International Symposium on Cluster, Cloud and Internet Computing}, CCGRID '20, 2020.

\bibitem{GPULifetimeSC20}
George Ostrouchov, Don Maxwell, Rizwan~A. Ashraf, Christian Engelmann, Mallikarjun Shankar, and James~H. Rogers.
\newblock Gpu lifetimes on titan supercomputer: Survival analysis and reliability.
\newblock In {\em Proceedings of the International Conference for High Performance Computing, Networking, Storage and Analysis}, SC '20, 2020.

\bibitem{InstructGPT}
Long Ouyang, Jeffrey Wu, Xu~Jiang, Diogo Almeida, Carroll Wainwright, Pamela Mishkin, Chong Zhang, Sandhini Agarwal, Katarina Slama, Alex Gray, John Schulman, Jacob Hilton, Fraser Kelton, Luke Miller, Maddie Simens, Amanda Askell, Peter Welinder, Paul Christiano, Jan Leike, and Ryan Lowe.
\newblock Training language models to follow instructions with human feedback.
\newblock In {\em Advances in Neural Information Processing Systems}, NeurIPS '22, 2022.

\bibitem{CarbonLLM}
David Patterson, Joseph Gonzalez, Quoc Le, Chen Liang, Lluis-Miquel Munguia, Daniel Rothchild, David So, Maud Texier, and Jeff Dean.
\newblock Carbon emissions and large neural network training.
\newblock {\em CoRR}, 2021.

\bibitem{Optimus}
Yanghua Peng, Yixin Bao, Yangrui Chen, Chuan Wu, and Chuanxiong Guo.
\newblock Optimus: An efficient dynamic resource scheduler for deep learning clusters.
\newblock In {\em Proceedings of the Thirteenth EuroSys Conference}, EuroSys '18, 2018.

\bibitem{Pollux}
Aurick Qiao, Sang~Keun Choe, Suhas~Jayaram Subramanya, Willie Neiswanger, Qirong Ho, Hao Zhang, Gregory~R. Ganger, and Eric~P. Xing.
\newblock Pollux: Co-adaptive cluster scheduling for goodput-optimized deep learning.
\newblock In {\em 15th {USENIX} Symposium on Operating Systems Design and Implementation}, OSDI '21, 2021.

\bibitem{Prometheus}
Bj{\"o}rn Rabenstein and Julius Volz.
\newblock Prometheus: A {Next-Generation} monitoring system (talk).
\newblock Dublin, 2015. USENIX Association.

\bibitem{CLIP}
Alec Radford, Jong~Wook Kim, Chris Hallacy, Aditya Ramesh, Gabriel Goh, Sandhini Agarwal, Girish Sastry, Amanda Askell, Pamela Mishkin, Jack Clark, Gretchen Krueger, and Ilya Sutskever.
\newblock Learning transferable visual models from natural language supervision.
\newblock In {\em Proceedings of the 38th International Conference on Machine Learning}, ICML '21, 2021.

\bibitem{GPT-1}
Alec Radford, Karthik Narasimhan, Tim Salimans, Ilya Sutskever, et~al.
\newblock Improving language understanding by generative pre-training.
\newblock 2018.

\bibitem{GPT-2}
Alec Radford, Jeff Wu, Rewon Child, David Luan, Dario Amodei, and Ilya Sutskever.
\newblock Language models are unsupervised multitask learners.
\newblock 2019.

\bibitem{ZeRO}
Samyam Rajbhandari, Jeff Rasley, Olatunji Ruwase, and Yuxiong He.
\newblock Zero: Memory optimizations toward training trillion parameter models.
\newblock In {\em Proceedings of the International Conference for High Performance Computing, Networking, Storage and Analysis}, SC '20, 2020.

\bibitem{ZeRO-infinity}
Samyam Rajbhandari, Olatunji Ruwase, Jeff Rasley, Shaden Smith, and Yuxiong He.
\newblock Zero-infinity: breaking the gpu memory wall for extreme scale deep learning.
\newblock In {\em Proceedings of the International Conference for High Performance Computing, Networking, Storage and Analysis}, SC '21. Association for Computing Machinery, 2021.

\bibitem{ZeRO-Offload}
Jie Ren, Samyam Rajbhandari, Reza~Yazdani Aminabadi, Olatunji Ruwase, Shuangyan Yang, Minjia Zhang, Dong Li, and Yuxiong He.
\newblock Zero-offload: Democratizing billion-scale model training.
\newblock In {\em 2021 {USENIX} Annual Technical Conference}, {USENIX} {ATC} '21, 2021.

\bibitem{Diffusion}
Robin Rombach, Andreas Blattmann, Dominik Lorenz, Patrick Esser, and Bj\"orn Ommer.
\newblock High-resolution image synthesis with latent diffusion models.
\newblock In {\em Proceedings of the IEEE/CVF Conference on Computer Vision and Pattern Recognition}, CVPR '22, 2022.

\bibitem{DistilBERT}
Victor Sanh, Lysandre Debut, Julien Chaumond, and Thomas Wolf.
\newblock Distilbert, a distilled version of bert: smaller, faster, cheaper and lighter.
\newblock {\em CoRR}, 2019.

\bibitem{MoE}
Noam Shazeer, *Azalia Mirhoseini, *Krzysztof Maziarz, Andy Davis, Quoc Le, Geoffrey Hinton, and Jeff Dean.
\newblock Outrageously large neural networks: The sparsely-gated mixture-of-experts layer.
\newblock In {\em International Conference on Learning Representations}, ICLR '17, 2017.

\bibitem{Megatron-LMV1}
Mohammad Shoeybi, Mostofa Patwary, Raul Puri, Patrick LeGresley, Jared Casper, and Bryan Catanzaro.
\newblock Megatron-lm: Training multi-billion parameter language models using model parallelism.
\newblock {\em CoRR}, 2020.

\bibitem{GPUerrorDSN21}
Amir Taherin, Tirthak Patel, Giorgis Georgakoudis, Ignacio Laguna, and Devesh Tiwari.
\newblock Examining failures and repairs on supercomputers with multi-gpu compute nodes.
\newblock In {\em IEEE/IFIP International Conference on Dependable Systems and Networks}, DSN '21, 2021.

\bibitem{NetBouncer}
Cheng Tan, Ze~Jin, Chuanxiong Guo, Tianrong Zhang, Haitao Wu, Karl Deng, Dongming Bi, and Dong Xiang.
\newblock {NetBouncer}: Active device and link failure localization in data center networks.
\newblock In {\em 16th USENIX Symposium on Networked Systems Design and Implementation}, NSDI '19, 2019.

\bibitem{Bamboo}
John Thorpe, Pengzhan Zhao, Jonathan Eyolfson, Yifan Qiao, Zhihao Jia, Minjia Zhang, Ravi Netravali, and Guoqing~Harry Xu.
\newblock Bamboo: Making preemptible instances resilient for affordable training of large dnns.
\newblock In {\em 20th {USENIX} Symposium on Networked Systems Design and Implementation}, NSDI '23, 2023.

\bibitem{GPUReliabilitySC15}
Devesh Tiwari, Saurabh Gupta, George Gallarno, Jim Rogers, and Don Maxwell.
\newblock Reliability lessons learned from gpu experience with the titan supercomputer at oak ridge leadership computing facility.
\newblock In {\em Proceedings of the International Conference for High Performance Computing, Networking, Storage and Analysis}, SC '15, 2015.

\bibitem{GPUerrorHPCA15}
Devesh Tiwari, Saurabh Gupta, James Rogers, Don Maxwell, Paolo Rech, Sudharshan Vazhkudai, Daniel Oliveira, Dave Londo, Nathan DeBardeleben, Philippe Navaux, Luigi Carro, and Arthur Bland.
\newblock Understanding gpu errors on large-scale hpc systems and the implications for system design and operation.
\newblock In {\em International Symposium on High Performance Computer Architecture}, HPCA '15, 2015.

\bibitem{LLaMA}
Hugo Touvron, Thibaut Lavril, Gautier Izacard, Xavier Martinet, Marie-Anne Lachaux, Timothée Lacroix, Baptiste Rozière, Naman Goyal, Eric Hambro, Faisal Azhar, Aurelien Rodriguez, Armand Joulin, Edouard Grave, and Guillaume Lample.
\newblock Llama: Open and efficient foundation language models.
\newblock {\em CoRR}, 2023.

\bibitem{LLaMA2}
Hugo Touvron, Louis Martin, Kevin Stone, Peter Albert, Amjad Almahairi, Yasmine Babaei, Nikolay Bashlykov, Soumya Batra, Prajjwal Bhargava, Shruti Bhosale, Dan Bikel, Lukas Blecher, Cristian~Canton Ferrer, Moya Chen, Guillem Cucurull, David Esiobu, Jude Fernandes, Jeremy Fu, Wenyin Fu, Brian Fuller, Cynthia Gao, Vedanuj Goswami, Naman Goyal, Anthony Hartshorn, Saghar Hosseini, Rui Hou, Hakan Inan, Marcin Kardas, Viktor Kerkez, Madian Khabsa, Isabel Kloumann, Artem Korenev, Punit~Singh Koura, Marie-Anne Lachaux, Thibaut Lavril, Jenya Lee, Diana Liskovich, Yinghai Lu, Yuning Mao, Xavier Martinet, Todor Mihaylov, Pushkar Mishra, Igor Molybog, Yixin Nie, Andrew Poulton, Jeremy Reizenstein, Rashi Rungta, Kalyan Saladi, Alan Schelten, Ruan Silva, Eric~Michael Smith, Ranjan Subramanian, Xiaoqing~Ellen Tan, Binh Tang, Ross Taylor, Adina Williams, Jian~Xiang Kuan, Puxin Xu, Zheng Yan, Iliyan Zarov, Yuchen Zhang, Angela Fan, Melanie Kambadur, Sharan Narang, Aurelien Rodriguez, Robert Stojnic, Sergey Edunov, and Thomas Scialom.
\newblock Llama 2: Open foundation and fine-tuned chat models.
\newblock {\em CoRR}, 2023.

\bibitem{Transformer}
Ashish Vaswani, Noam Shazeer, Niki Parmar, Jakob Uszkoreit, Llion Jones, Aidan~N Gomez, \L~ukasz Kaiser, and Illia Polosukhin.
\newblock Attention is all you need.
\newblock In {\em Advances in Neural Information Processing Systems}, NeurIPS '17, 2017.

\bibitem{HFTA}
Shang Wang, Peiming Yang, Yuxuan Zheng, Xin Li, and Gennady Pekhimenko.
\newblock Horizontally fused training array: An effective hardware utilization squeezer for training novel deep learning models.
\newblock In {\em Proceedings of Machine Learning and Systems}, MLSys '21, 2021.

\bibitem{SelfConsistency}
Xuezhi Wang, Jason Wei, Dale Schuurmans, Quoc~V Le, Ed~H. Chi, Sharan Narang, Aakanksha Chowdhery, and Denny Zhou.
\newblock Self-consistency improves chain of thought reasoning in language models.
\newblock In {\em International Conference on Learning Representations}, ICLR '23, 2023.

\bibitem{GEMINI}
Zhuang Wang, Zhen Jia, Shuai Zheng, Zhen Zhang, Xinwei Fu, T.~S.~Eugene Ng, and Yida Wang.
\newblock Gemini: Fast failure recovery in distributed training with in-memory checkpoints.
\newblock In {\em Proceedings of the ACM SIGOPS 29th Symposium on Operating Systems Principles}, SOSP '23, 2023.

\bibitem{MLaaS}
Qizhen Weng, Wencong Xiao, Yinghao Yu, Wei Wang, Cheng Wang, Jian He, Yong Li, Liping Zhang, Wei Lin, and Yu~Ding.
\newblock {MLaaS} in the wild: Workload analysis and scheduling in {Large-Scale} heterogeneous {GPU} clusters.
\newblock In {\em 19th USENIX Symposium on Networked Systems Design and Implementation}, NSDI '22, 2022.

\bibitem{Gandiva}
Wencong Xiao, Romil Bhardwaj, Ramachandran Ramjee, Muthian Sivathanu, Nipun Kwatra, Zhenhua Han, Pratyush Patel, Xuan Peng, Hanyu Zhao, Quanlu Zhang, Fan Yang, and Lidong Zhou.
\newblock Gandiva: Introspective cluster scheduling for deep learning.
\newblock In {\em 13th {USENIX} Symposium on Operating Systems Design and Implementation}, OSDI '18, 2018.

\bibitem{AntMan}
Wencong Xiao, Shiru Ren, Yong Li, Yang Zhang, Pengyang Hou, Zhi Li, Yihui Feng, Wei Lin, and Yangqing Jia.
\newblock Antman: Dynamic scaling on {GPU} clusters for deep learning.
\newblock In {\em 14th {USENIX} Symposium on Operating Systems Design and Implementation}, OSDI '20, 2020.

\bibitem{Anubis}
Yifan Xiong, Yuting Jiang, Ziyue Yang, Lei Qu, Guoshuai Zhao, Shuguang Liu, Dong Zhong, Boris Pinzur, Jie Zhang, Yang Wang, Jithin Jose, Hossein Pourreza, Jeff Baxter, Kushal Datta, Prabhat Ram, Luke Melton, Joe Chau, Peng Cheng, Yongqiang Xiong, and Lidong Zhou.
\newblock Anubis: Towards reliable cloud ai infrastructure via proactive validation.
\newblock {\em CoRR}, abs/2402.06194, 2024.

\bibitem{ASTRAEA}
Zhisheng Ye, Peng Sun, Wei Gao, Tianwei Zhang, Xiaolin Wang, Shengen Yan, and Yingwei Luo.
\newblock Astraea: A fair deep learning scheduler for multi-tenant gpu clusters.
\newblock {\em IEEE Transactions on Parallel and Distributed Systems}, 2021.

\bibitem{SLURM}
Andy~B. Yoo, Morris~A. Jette, and Mark Grondona.
\newblock Slurm: Simple linux utility for resource management.
\newblock In {\em Job Scheduling Strategies for Parallel Processing}, 2003.

\bibitem{Zeus}
Jie You, Jae-Won Chung, and Mosharaf Chowdhury.
\newblock Zeus: Understanding and optimizing {GPU} energy consumption of {DNN} training.
\newblock In {\em 20th USENIX Symposium on Networked Systems Design and Implementation}, NSDI '23, 2023.

\bibitem{Orca}
Gyeong-In Yu, Joo~Seong Jeong, Geon-Woo Kim, Soojeong Kim, and Byung-Gon Chun.
\newblock Orca: A distributed serving system for {Transformer-Based} generative models.
\newblock In {\em 16th USENIX Symposium on Operating Systems Design and Implementation}, OSDI '22, 2022.

\bibitem{SNAP}
Minlan Yu, Albert Greenberg, Dave Maltz, Jennifer Rexford, Lihua Yuan, Srikanth Kandula, and Changhoon Kim.
\newblock Profiling network performance for multi-tier data center applications.
\newblock In {\em 8th USENIX Symposium on Networked Systems Design and Implementation}, NSDI '11, 2011.

\bibitem{Salus}
Peifeng Yu and Mosharaf Chowdhury.
\newblock Fine-grained gpu sharing primitives for deep learning applications.
\newblock In {\em Proceedings of Machine Learning and Systems}, MLSys '20, 2020.

\bibitem{SHEPHERD}
Hong Zhang, Yupeng Tang, Anurag Khandelwal, and Ion Stoica.
\newblock {SHEPHERD}: Serving {DNNs} in the wild.
\newblock In {\em 20th USENIX Symposium on Networked Systems Design and Implementation}, NSDI '23, 2023.

\bibitem{Deepview}
Qiao Zhang, Guo Yu, Chuanxiong Guo, Yingnong Dang, Nick Swanson, Xinsheng Yang, Randolph Yao, Murali Chintalapati, Arvind Krishnamurthy, and Thomas Anderson.
\newblock Deepview: Virtual disk failure diagnosis and pattern detection for azure.
\newblock In {\em 15th USENIX Symposium on Networked Systems Design and Implementation}, NSDI '18, 2018.

\bibitem{DLFailures}
Ru~Zhang, Wencong Xiao, Hongyu Zhang, Yu~Liu, Haoxiang Lin, and Mao Yang.
\newblock An empirical study on program failures of deep learning jobs.
\newblock In {\em Proceedings of the ACM/IEEE 42nd International Conference on Software Engineering}, ICSE '20, 2020.

\bibitem{OPT}
Susan Zhang, Stephen Roller, Naman Goyal, Mikel Artetxe, Moya Chen, Shuohui Chen, Christopher Dewan, Mona Diab, Xian Li, Xi~Victoria Lin, Todor Mihaylov, Myle Ott, Sam Shleifer, Kurt Shuster, Daniel Simig, Punit~Singh Koura, Anjali Sridhar, Tianlu Wang, and Luke Zettlemoyer.
\newblock Opt: Open pre-trained transformer language models.
\newblock {\em CoRR}, 2022.

\bibitem{Justitia}
Yiwen Zhang, Yue Tan, Brent Stephens, and Mosharaf Chowdhury.
\newblock Justitia: Software {Multi-Tenancy} in hardware {Kernel-Bypass} networks.
\newblock In {\em 19th USENIX Symposium on Networked Systems Design and Implementation}, NSDI '22, 2022.

\bibitem{Arena}
Lianmin Zheng, Wei-Lin Chiang, Ying Sheng, Siyuan Zhuang, Zhanghao Wu, Yonghao Zhuang, Zi~Lin, Zhuohan Li, Dacheng Li, Eric.~P Xing, Hao Zhang, Joseph~E. Gonzalez, and Ion Stoica.
\newblock Judging llm-as-a-judge with mt-bench and chatbot arena.
\newblock {\em CoRR}, 2023.

\bibitem{Alpa}
Lianmin Zheng, Zhuohan Li, Hao Zhang, Yonghao Zhuang, Zhifeng Chen, Yanping Huang, Yida Wang, Yuanzhong Xu, Danyang Zhuo, Eric~P. Xing, Joseph~E. Gonzalez, and Ion Stoica.
\newblock Alpa: Automating inter- and {Intra-Operator} parallelism for distributed deep learning.
\newblock In {\em 16th USENIX Symposium on Operating Systems Design and Implementation}, OSDI '22, 2022.

\end{thebibliography}
